\documentclass{aa}  

\usepackage{graphicx}
\usepackage{txfonts}
\usepackage{natbib,twoopt}
\usepackage{longtable}

\begin{document}

   \title{Rubidium and zirconium abundances in massive Galactic asymptotic giant branch stars revisited}

   \author{V. P\'erez-Mesa\inst{1,2}, O. Zamora\inst{1,2}, D. A.
   Garc\'ia-Hern\'andez\inst{1,2}, B. Plez\inst{3}, A. Manchado\inst{1,2,4}, A. I. Karakas\inst{5}
          \and
          M. Lugaro\inst{5,6}
          }

   \institute{Instituto de Astrof\'isica de Canarias (IAC), E-38205 La Laguna, Tenerife, Spain\\
              \email{vperezme@iac}
         \and
             Departamento de Astrof\'isica, Universidad de La Laguna (ULL), 
             E-38206 La Laguna, Tenerife, Spain
         \and
             Laboratoire Univers et Particules de Montpellier, Universit\'e 
             de Montpellier2, CNRS, 34095 Montpellier, France
             \and
             Consejo Superior de Investigaciones Cient\'ificas (CSIC), E-28006 
             Madrid, Spain
             \and
              Monash Centre for Astrophysics, School of Physics and Astronomy, 
              Monash University, VIC3800, Australia
             \and
             Konkoly Observatory, Research Centre for Astronomy and Earth Sciences, 
             Hungarian Academy of Sciences, 1121 Budapest, Hungary\\
             }

   \date{Received September 15, 1996; accepted March 16, 1997}

\titlerunning{Rb and Zr abundances in massive Galactic AGB stars revisited}
\authorrunning{V. P\'erez-Mesa et al.}

 
  \abstract
   {Luminous Galactic OH/IR stars have been identified as massive
($>$ 4-5 M$_{\odot}$) asymptotic giant branch (AGB) stars experiencing hot
bottom burning and Li production. Their Rb abundances and [Rb/Zr] ratios, as
derived from classical hydrostatic model atmospheres, are significantly higher
than predictions from AGB nucleosynthesis models, posing a problem to our 
understanding of AGB evolution and nucleosynthesis.}
   {We report new Rb and Zr abundances in the full sample (21) of massive 
   Galactic AGB stars,
previously studied with hydrostatic models, by using more realistic extended
model atmospheres.}
   {For this, we use a modified version of the spectral synthesis code
Turbospectrum and consider the presence of a circumstellar envelope and radial
wind in the modelling of the optical spectra of these massive AGB stars. The Rb
and Zr abundances are determined from the 7800 {\AA} Rb I resonant line and the
6474 {\AA} ZrO bandhead, respectively, and we explore the sensitivity of the
derived abundances to variations of the stellar (T$_{eff}$) and wind ($\dot{M}$,
$\beta$ and $v_{exp}$) parameters in the pseudo-dynamical models. The Rb and Zr
abundances derived from the best spectral fits are compared with the most
recent AGB nucleosynthesis theoretical predictions.}
   {The Rb abundances derived with the pseudo-dynamical models are much lower 
(in the most extreme stars even by $\sim$1-2 dex) than those derived with the 
hydrostatic models, while the Zr abundances are similar. The Rb I line profile 
and Rb abundance are very sensitive to the wind mass-loss rate $\dot{M}$ 
(especially for $\dot{M}$ $\geq$ 10$^{-8}$ M$_{\odot}yr^{-1}$) but much less 
sensitive to variations of the wind velocity-law ($\beta$ parameter) and the 
expansion velocity $v_{exp}$(OH).}
   {We confirm the earlier preliminary results based on a smaller sample of 
massive O-rich AGB stars, that the use of extended atmosphere models can 
solve the discrepancy
between the AGB nucleosynthesis theoretical models and the observations of
Galactic massive AGB stars. The Rb abundances, however, are still strongly
dependent of the wind mass-loss $\dot{M}$, which, unfortunately, is unknown in
these AGB stars. Accurate mass-loss rates $\dot{M}$ (e.g., from rotationally
excited lines of the CO isotopologues in the radio domain) in these massive
Galactic AGB stars are needed in order to break the models degeneracy and get
reliable (no model-dependent) Rb abundances in these stars.}

   \keywords{stars: AGB and post-AGB --
                stars: abundances --
                stars: evolution --
                nuclear reactions, nucleosynthesis, abundances --
                stars: atmospheres --
                stars: late-type
               }

   \maketitle
%

\section{Introduction}

The asymptotic giant brach (AGB; \citealt{herwig05, karakaslattanzio14}) 
is occupied by low- and
intermediate-mass (0.8 $\leq$ M $\leq$ 8 M$_{\odot}$) stars in the last
nuclear-burning phase. At the end of the AGB phase, these stars develop 
thermal pulses (TP) and suffer extreme mass loss. AGB stars are thus 
one of the main contributors to the enrichment of the
interstellar medium (ISM) of light elements (e.g. Li, C, N, F) and heavy
(\textit{slow} neutron capture, \textit{s}-process) elements and so to the 
chemical evolution of galaxies \citep{busso99}. AGB stars are also one of 
the most prominent source of dust in galaxies and the site of origin of the 
vast majority of meteoritic stardust grains \citep[e.g.][]{hoppeott97, nittler97,
lugaro17}. In low-mass AGB stars (M $<$ 4 M$_{\odot}$) $^{12}$C is produced 
during the TP-AGB phase, and carried to the stellar surface via the third 
dredge-up (TDU) that can occur after each TP, transforming originally O-rich 
stars in C-rich stars (C/O$>$1) \citep[e.g.][]{herwig05, karakas07, lugaro11}. 
However, the more massive AGB stars (M $>$ 4-5 M$_{\odot}$) are O-rich (C/O$<$1) 
because the so-called "hot bottom burning" (hereafter, HBB) process is activated. 
The HBB converts $^{12}$C into $^{13}$C and $^{14}$N through the CN cycle via proton 
captures at the base of the convective envelope, thus preventing the formation of 
a carbon star \citep{sackmann92, mazzitelli99}.  

The \textit{s}-process allows the production of neutron-rich elements heavier than
iron (\textit{s}-elements such as Sr, Y, Zr, Ba, La, Nd, Tc, etc.) by \textit{s}
-process. In the low-mass AGB stars (roughly < 4 M$_{\odot}$), the $^{13}$C($\alpha$, 
n)$^{16}$O reaction is the dominant neutron source \citep[e.g.][]{abia01}. In the 
more massive AGB stars instead, neutrons are mainly released by the $^{22}$Ne($\alpha$, 
n)$^{25}$Mg reaction, resulting in a higher neutron density (up to 10$^{13}$ n/cm$^{3}$)
and temperature environment than in lower mass AGB stars \citep{garcia-hernandez06}.
The Rb produced depends on the probability of the $^{85}$Kr and $^{86}$Rb capturing a
neutron before decaying and acting as "branching points" \citep[see][for more 
details]{vanraai12}. The probabiliy of this happening depends on the local neutron density
\citep{beermacklin89}. The  $^{87}$Rb/$^{85}$Rb isotopic ratio is a direct indicator
of the neutron density at the production site but it is not possible to distinguish
individual $^{87}$Rb and $^{85}$Rb from stellar spectra \citep{garcia-hernandez06}.
However, the relative abundance of Rb to other nearby \textit{s}-process elements such 
as Zr is very sensitive to the neutron density, and so a good discriminant of the stellar
mass and the neutron source at the $s$-process site \citep{lambert95, abia01, 
garcia-hernandez06, vanraai12}. In other words, [Rb/Zr]$<$0 is observed in low-mass
AGB stars where the main neutron source is the $^{13}$C($\alpha$, n)$^{16}$O
reaction \citep{plez93, lambert95, abia01}, while [Rb/Zr]$>$0 is observed in more
massive AGB stars, where the neutrons are mainly  released through the
$^{22}$Ne($\alpha$, n)$^{25}$Mg reaction \citep{garcia-hernandez06,
garcia-hernandez07, garcia-hernandez09}. 

Chemical abundance analyses using classical MARCS hydrostatic atmospheres
\citep{gustafsson08} revealed strong Rb overabundances ($\sim$10$^{3}$-10$^{5}$
times solar) and high [Rb/Zr] ratios ($\geqslant$ 3-4 dex) in massive AGB stars 
(generally very luminous OH/IR stars) of our own Galaxy and the Magellanic Clouds 
(MC; \citealt{garcia-hernandez06, garcia-hernandez07, garcia-hernandez09}). This 
observationally confirmed for the first time that the $^{22}$Ne neutron source dominates 
the production of \textit{s}-process elements in these stars. However, the extremely 
high Rb abundances and [Rb/Zr] ratios observed in most the massive stars (and especially 
in the lower metallicity MC AGB stars) have posed a "Rb problem"; such extreme [Rb/Fe] 
and [Rb/Zr] values are not predicted by the \textit{s}-process AGB models, 
\citep[][]{vanraai12, karakas12}, suggesting fundamental problems in our present 
understanding of AGB nucleosynthesis and/or of the complex extended dynamical atmospheres of 
these stars \citep{garcia-hernandez09}. 

\cite{zamora14} constructed new pseudo-dynamical MARCS model atmospheres by considering
the presence of a gaseous circumstellar envelope with a radial wind and
applied them to a small sample of five O-rich AGB stars with different expansion
velocities and metallicities. The Rb abundances and [Rb/Zr] ratios obtained were
much lower than those obtained with classical hydrostatic models; in better
agreement with the AGB nucleosynthesis theoretical predictions. In this paper,
we use the \cite{zamora14} pseudo-dynamical model atmospheres to obtain the abundances
of Rb and Zr in the full sample of massive Galactic AGB stars previously
analyzed with hydrostatic models \citep{garcia-hernandez06, garcia-hernandez07}.
These Rb and Zr abundances are then compared with the more recent AGB
nucleosynthesis theoretical predictions available in the literature.


\section{Sample and observational data}

Our sample is composed by 21 massive Galactic AGB stars (most of them very
luminous OH/IR stars) previously analyzed by \citet{garcia-hernandez06,
garcia-hernandez07}; we use their high-resolution ($R\sim$40,000$-$50,000) 
optical echelle spectra \citep[see][for further observational details]
{garcia-hernandez06,garcia-hernandez07}\footnote{The high-resolution spectra 
were obtained using the Utrecht Echelle Spectrograph (UES) at the 4.2 m William 
Herschel Telescope (La Palma, Spain) and the CAsegrain Echelle SPECtrograph 
(CASPEC) of the ESO 3.6 m telescope (La Silla, Chile) during several observing 
periods in 1996-97 \citep[see][]{garcia-hernandez07}.}. The signal-to-noise ($S/N$)
ratios achieved in the reduced spectra strongly vary from the blue to the red
(typically $\sim$ 10-20 at 6000 $\AA$ and $>$100 at 8000 \AA). The Rb and
Zr abundances were determined from the resonant 7800 $\AA$ Rb I line and the
6474 $\AA$ ZrO bandhead, respectively, by using classical MARCS hydrostatic
model atmospheres \citep{garcia-hernandez06,garcia-hernandez07}. The Rb
abundances and [Rb/Zr] ratios obtained from this chemical analysis are mostly in
the range [Rb/Fe]$\sim$0.6$-$2.6 dex and [Rb/Zr]$\sim$0.1$-$2.1 dex. The
atmospheric parameters and Rb abundances derived with the hydrostatic models as
well as other useful observational information like the OH expansion velocity,
variability period, and the presence of Li are listed in Table~\ref{table_obs_param}.

\begin{table*}
\centering
\caption{Atmosphere parameters and Rb abundances (as derived using hydrostatic
models) and other selected observational information. \label{table_obs_param}}
\renewcommand{\arraystretch}{1.25}
\begin{tabular}{ccccccccc}
\hline
\hline

IRAS name  & $T_{eff}$ (K) & log $g$ & $v_{exp}$(OH) (km s$^{-1}$)& Period (days) & Lithium & {[}Rb/Fe{]}$_{static}$ & S/N at 7800 \AA \\

\hline

01085$+$3022 & 3000$^{*}$ & $-$0.5 & 13   & 560 & yes  & 2.0     & 49  \\
04404$-$7427 & 3000       & $-$0.5 & 8    & 534 & ...  & 1.3     & 68  \\
05027$-$2158 & 2800       & $-$0.5 & 8    & 368 & yes  & 0.4     & 418 \\
05098$-$6422 & 3000       & $-$0.5 & 6    & 394 & no   & 0.1     & 309 \\
05151$+$6312 & 3000       & $-$0.5 & 15   & ... & no   & 2.1     & 161 \\
06300$+$6058 & 3000       & $-$0.5 & 12   & 440 & yes  & 1.6     & 127 \\
07222$-$2005 & 3000       & $-$0.5 & 8    & 1200& ...  & 0.6     & 30  \\
09194$-$4518 & 3000       & $-$0.5 & 11   & ... & ...  & 1.1     & 25  \\
10261$-$5055 & 3000       & $-$0.5 & 4    & 317 & no   & $<-$1.0 & 595 \\
14266$-$4211 & 2900       & $-$0.5 & 9    & 389 & no   & 0.9     & 106 \\
15193$+$3132 & 2800       & $-$0.5 & 3    & 360 & no   & $-$0.3    & 266 \\
15576$-$1212 & 3000       & $-$0.5 & 10   & 415 & yes  & 1.5     & 91  \\
16030$-$5156 & 3000       & $-$0.5 & 7-14 & 579 & yes  & 1.3     & 86  \\
16037$+$4218 & 2900       & $-$0.5 & 4    & 360 & no   & 0.6     & 115 \\
17034$-$1024 & 3300       & $-$0.5 & 3-9  & 346 & no   & 0.2     & 189 \\
18429$-$1721 & 3000       & $-$0.5 & 7    & 481 & yes  & 1.2     & 98   \\
19059$-$2219 & 3000       & $-$0.5 & 13   & 510 & ...  & 2.3     & 32   \\
19426$+$4342 & 3000       & $-$0.5 & 9    & ... & ...  & 1.0     & 19   \\
20052$+$0554 & 3000$^{*}$ & $-$0.5 & 16   & 450 & yes  & 1.5     & 47   \\
20077$-$0625 & 3000       & $-$0.5 & 12   & 680 & ...  & 1.3     & 19   \\
20343$-$3020 & 3000       & $-$0.5 & 8    & 349 & no   & 0.9     & 76   \\

\hline  
\end{tabular}
\tablefoot{The stellar parameters, Rb abundances, OH expansion velocities,
variability periods and presence of Li are collected from
\citet{garcia-hernandez06,garcia-hernandez07} (and references therein). The
asterisks indicate that the best fitting $T_{eff}$ in the ZrO 6474 $\AA$
spectral region is warmer (3300 K) than that around the Rb I 7800 $\AA$ line
\citep{garcia-hernandez06,garcia-hernandez07}. Two stars (IRAS 16030$-$5156 and
IRAS 17034$-$1024) only display the blue-shifted 1612 MHz OH maser peak and we
list the range of OH expansion velocities shown by other stars with similar
variability periods \citep{garcia-hernandez07}.}
\end{table*}


\section{Chemical abundance analysis using pseudo-dynamical models}

\subsection{Modified version of the Turbospectrum spectral synthesis code}

We have used the v12.2 version of the spectral synthesis code
\textit{Turbospectrum} \citep{alvarez98, plez12}, which considers the presence
of a circumstellar gas envelope and a radial wind, as modified by \cite{zamora14}.
The main modifications are the following: (i) the Doppler effect due to the
extended atmosphere and velocity field is introduced in the routines that
compute the line intensities at the stellar surface; (ii) the source
function of the radiative transfer is assumed to be the same as computed in the
static case \citep{gustafsson08}. The validity of this approximation was tested
by comparing with Monte Carlo simulations \citep[see][]{zamora14}; (iii) 
the scattering term
of the source function ($\varpropto\sigma_{\lambda}J_{\lambda}$) is not shifted
to save computing time and it is only incorporated for the continuum.
This scattering term is computed as in the static case
using the Feautrier method \citep{nordlund84, gustafsson08}; and (iv) the velocity field
is taken into account through a shift of the absorption coefficient
$\kappa_{\lambda}$; the source function is built using the static
$\sigma_{\lambda}J_{\lambda}$ and the shifted $\kappa_{\lambda}B_{\lambda}$. 
The emerging intensity is then computed in the observer frame by a direct quadrature 
of the source function.

\subsection{Extended atmosphere models}

For the analysis of each star in our sample, we have adopted the atmosphere
parameters from \citet{garcia-hernandez06, garcia-hernandez07} and the solar
reference abundances by \cite{grevesse07}. We constructed our pseudo-dynamical models
from the original MARCS hydrostatic atmosphere model structure. We expanded the
atmosphere radius by a wind out to $\sim$5 stellar radii and a radial velocity
field \citep{zamora14}. In the MARCS hydrostatic model, the $R_{\ast}$ is the
radius corresponding to $r(\tau_{Ross}=1)$, where $r$ is the distance from the
center of the star and $\tau_{Ross}$ is the Rosseland optical depth. We have
computed the stellar wind following the mass conservation (Eq. 1), radiative
thermal equilibrium (Eq. 2) and a classical $\beta$-velocity law (Eq. 3),

	\begin{equation}
		\rho(r) = \frac{{\textit{\.M}}}{4 \pi r^2 v(r)} 
	\end{equation}

	\begin{equation}
		r T^2 = constant  = r_{out}T^2_{out}
	\end{equation}

	\begin{equation}
		v(r) = v_0+(v_{\infty}-v_0)\left(1-\frac{R_*}{r}\right)^{\beta} \,,
	\end{equation}

\noindent 
where $\rho(r)$ is the density of the envelope radius $r$, $\dot{M}$ is the
mass-loss rate and $v(r)$ is the velocity of the envelope, which is calculated
by means of Eq.(3). In Eq.(3), $v_{0}$ is a reference velocity for the
beginning of the wind and $\beta$ is an arbitrary free parameter. We take
$v_{0}=v(R_{\ast})$ for the onset of the wind and the extension of the envelope
begins from the outer radius of the hydrostatic model. Using Eq.(2) the envelope
is extended, layer by layer, out to the distance $r_{max}$, which corresponds to
the maximum radius in our calculations, with $T_{min}=1000$ K.
\textit{Turbospectrum} cannot compute lower temperatures due to numerical
reasons \citep{zamora14}. 

\begin{figure}
\centering
\includegraphics[width=9.3cm]{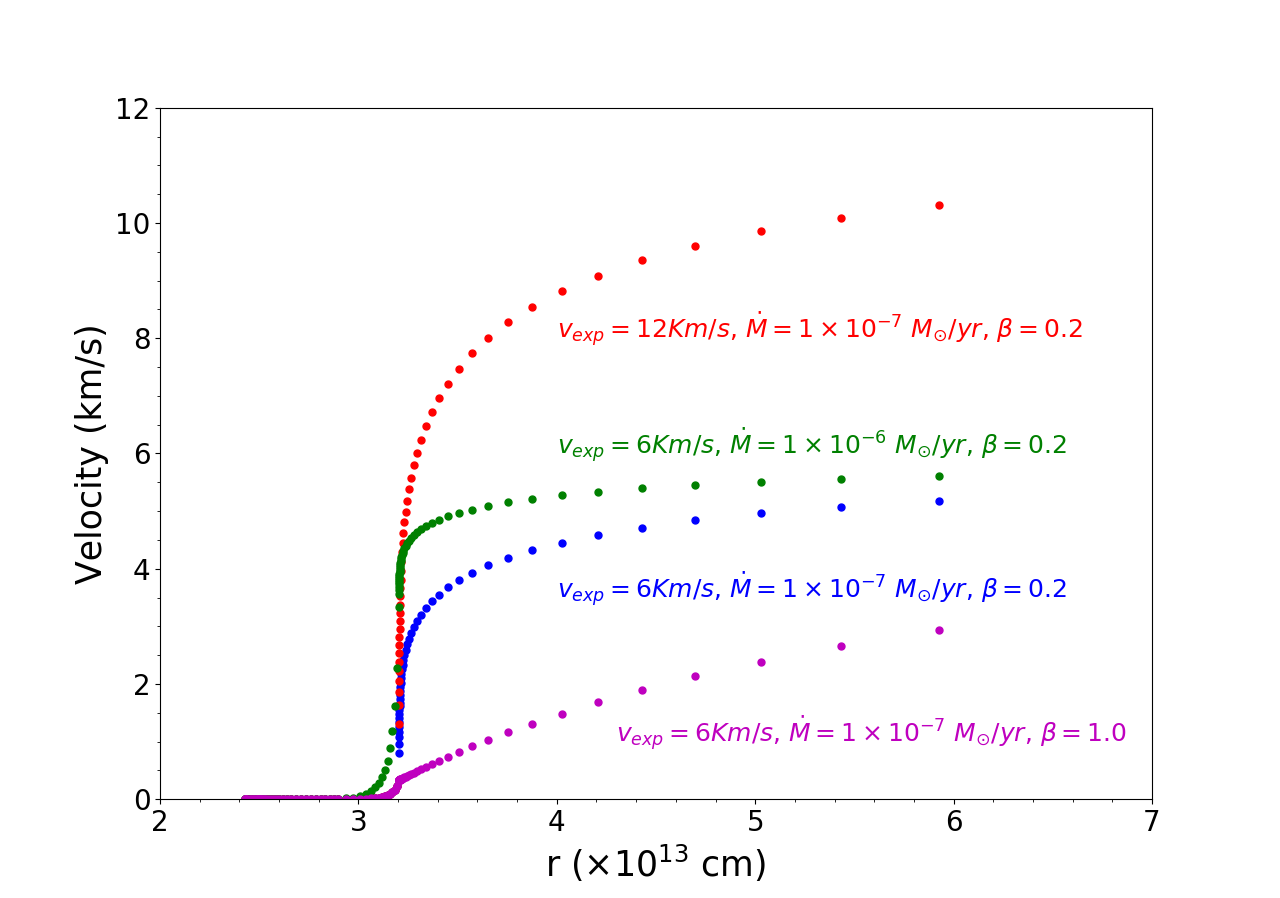}
\caption{Velocity vs. distance from the star in four of our AGB wind models.
These velocity laws present different expansion velocities $v_{exp}$(OH),
mass-loss rates $\dot{M}$ and $\beta$ exponents. The effective temperature
$T_{eff}$ = 3000 K, gravity log $g$ = $-$0.5 and the solar chemical composition
are the same in all models.}
\label{v_r}
\end{figure}

\subsection{Resulting grids of synthetic spectra}

The synthetic spectra are generated with the modified version of
\textit{Turbospectrum} by using the extended pseudo-dynamical model atmospheres as
input. We constructed a mini-grid of synthetic spectra for each sample star by
adopting the atmospheric parameters (e.g., effective temperature,
macroturbulence\footnote{The synthetic spectra are convolved with a Gaussian
profile (with a certain FWHM typically between 250 and 400 m\AA) to account for
macroturbulence as well as instrumental profile effects.}) from
\citet{garcia-hernandez06, garcia-hernandez07}. Basically, the stellar mass,
gravity log $g$, microturbulent velocity $\xi$, metallicity [Fe/H], and C/O ratio
are fixed to 2 M$_{\odot}$, $-$0.5 dex, 3 kms$^{-1}$, 0.0, and 0.5 dex,
respectively \citep[see][for more details]{garcia-hernandez07}. On the other
hand, for the mass-loss rate $\dot{M}$ and the exponent $\beta$, we use values
between $\dot{M} \sim 10^{-9}-10^{-6} M_{\odot}yr^{-1}$ in steps of
$0.5\times10^{-1}$ $M_{\odot}yr^{-1}$ and $\beta \sim 0.2-1.6$ in steps of 0.2.
We have not considered the case where $\beta = 0.0$ because the expansion
velocity would be constant at any $r$. We assume the OH expansion velocity
($v_{exp}$(OH); see Table~\ref{table_obs_param}) as the terminal velocity
because the OH maser emission is found at very large distances of
the central star \citep[see e.g.,][]{decin10}. Figure \ref{v_r} shows examples
of the $\beta$-velocity laws used in our pseudo-dynamical models based on the MARCS
hydrostatic models. Finally, for the Rb and Zr abundances we used
[Rb/Fe]$\sim-$2.6 to $+$3.0 dex, and [Zr/Fe]$\sim-$1.0 to $+$1.0 in steps of 0.1
and 0.25 dex, respectively.

The resulting mini-grid ($\sim$4400 models) is compared to the observed
spectrum in order to find the synthetic spectrum that best fits the 7800 $\AA$
Rb I line and the 6474 $\AA$ ZrO bandhead profiles and their adjacent
pseudocontinua. In order to obtain the best fits, we made use of a procedure
based on the comparison between synthetic and observed spectra, while in 
\cite{zamora14} the observed spectra were fitted by eye. The method is a
modified version of the standard $\chi^{2}$ test,

	\begin{equation}
		\chi^{2*}=\chi^{2} \times w = \left(\sum_{i=1}^{N}\dfrac{[Yobs_{i}-Ysynth_{i}
		(x_{1}...x_{M})]^2}{Y obs_{i}} \right) \times w
	\end{equation}

\noindent
where $Y obs_{i}$ and $Ysynth_{i}$ are the observed and synthetic data points,
respectively, with $N$ the number of data points, and $M$ the number of free
parameters. On the other hand, $w$ is a vector that gives a stronger weight to
the detailed spectral profiles of the Rb I line and the ZrO bandhead. This
way, the lowest value of $\chi^{2*}$ gives us the best fitting synthetic
spectrum from the mini-grid for each sample star. 

The use of the $\chi^{2*}$ test to find the best fits to the observed
spectra
reveals the presence of important
degeneracies in the resulting grids of pseudo-dynamical synthetic spectra; i.e., very
similar synthetic spectra are obtained from different sets of wind parameters
(see below for more details). Moreover, in a some cases (IRAS 04404$-$7427,
IRAS 05027$-$2158, IRAS 05098$-$6422, IRAS 06300$+$6058, IRAS 10261$-$5055, 
IRAS 18429$-$1721, IRAS 19059$-$2219 and 
IRAS 20343$-$3020) the use of the $\chi^{2*}$ test is not enough for obtaining the 
synthetic
spectrum that best reproduces the observed one and the best fits have to be
found by eye. Unfortunately, the wind model parameters $\dot{M}$ and $\beta$ are
generally not known for stars in our sample (see below), complicating the
abundance analysis. Thus, here we study the sensitivity of the synthetic spectra
and the abundance results to variations of the stellar and wind parameters. 

\subsection{Sensitivity of the synthetic spectra to variations of the model
parameters}\label{sec:variationofparameters}

\begin{figure*}
   \centering
   \includegraphics[width=9.15cm,height=6.7cm,angle=0]{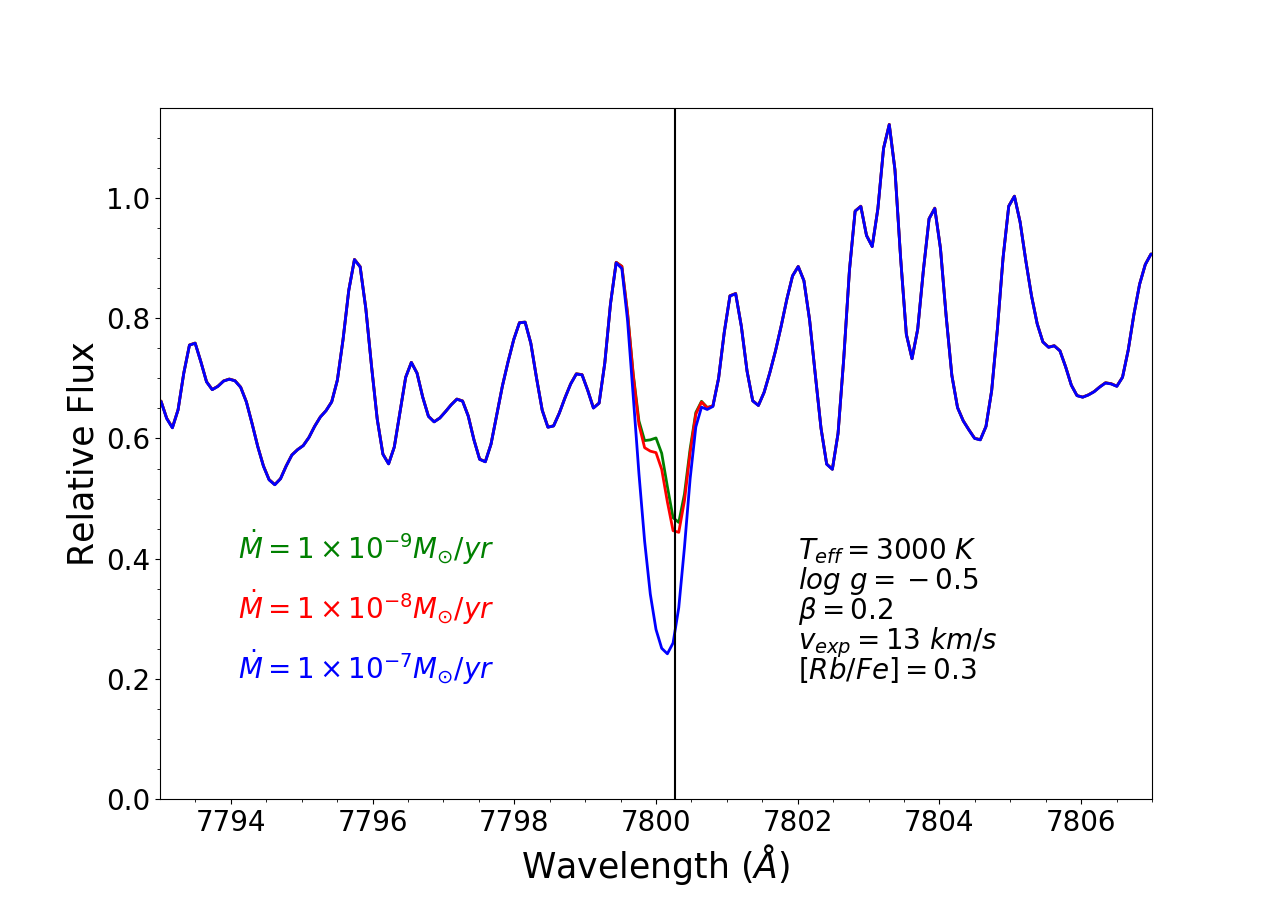}
   \includegraphics[width=9.15cm,height=6.7cm,angle=0]{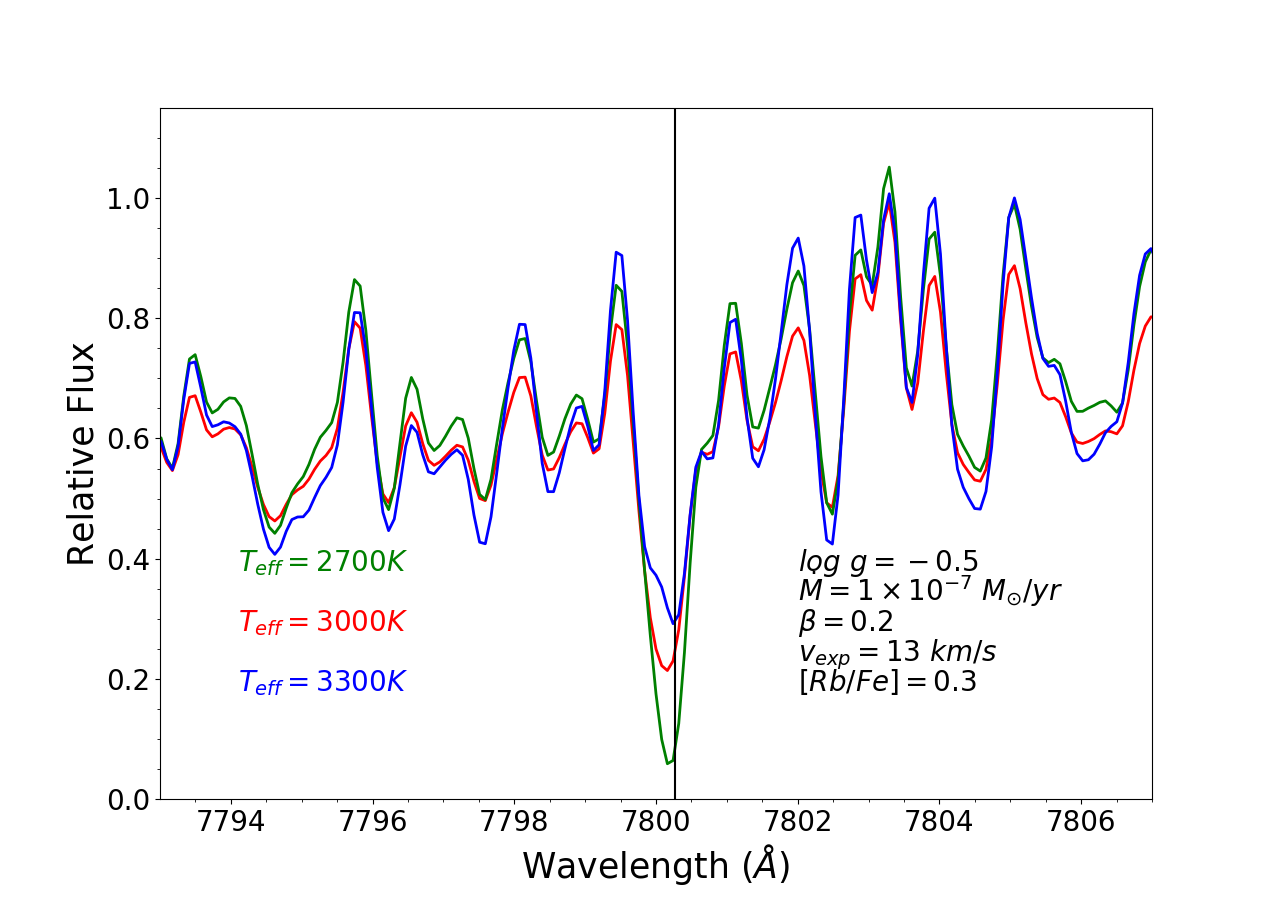}
   \includegraphics[width=9.15cm,height=6.7cm,angle=0]{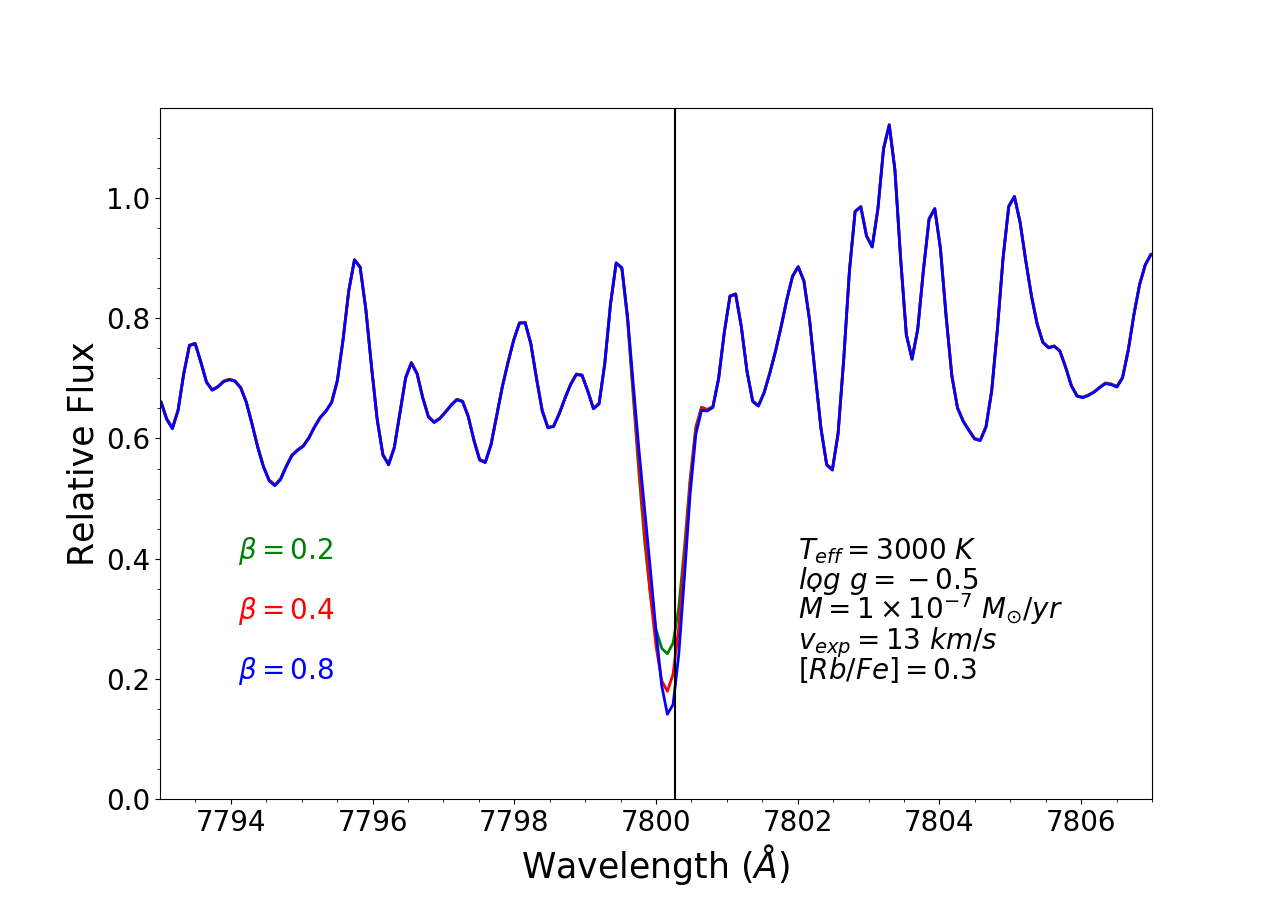}
   \includegraphics[width=9.15cm,height=6.7cm,angle=0]{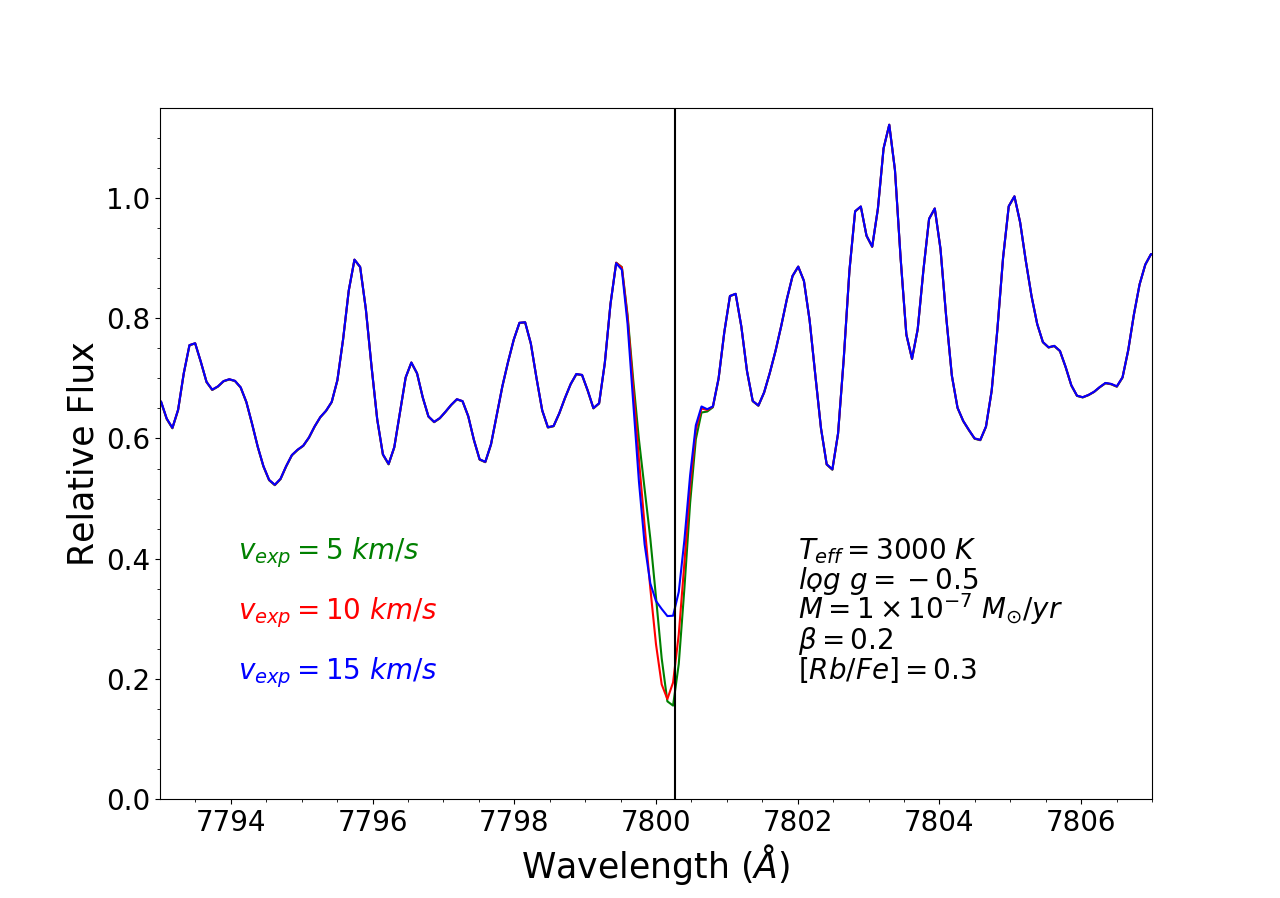}
      \caption{Illustrative examples of synthetic spectra for different stellar
(T$_{eff}$) and wind ($\dot{M}$, $\beta$, and $v_{exp}$(OH)) parameters in the
spectral region around the 7800 $\AA$ Rb I line. The black vertical line
indicates the position of the 7800 $\AA$ Rb I line.}
         \label{comparisons_Rb}
   \end{figure*}

\begin{figure*}
   \centering
   \includegraphics[width=9.15cm,height=6.7cm,angle=0]{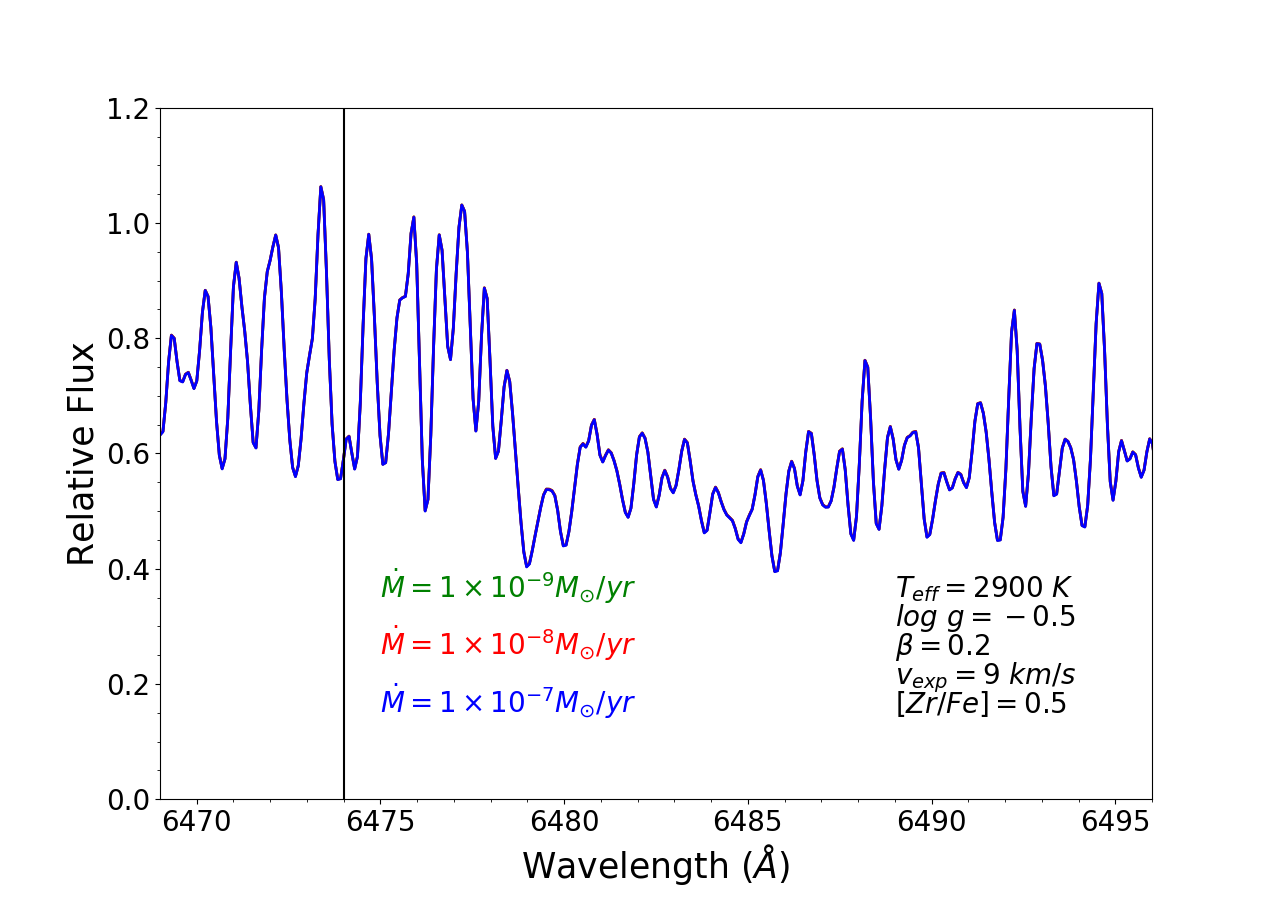}
   \includegraphics[width=9.15cm,height=6.7cm,angle=0]{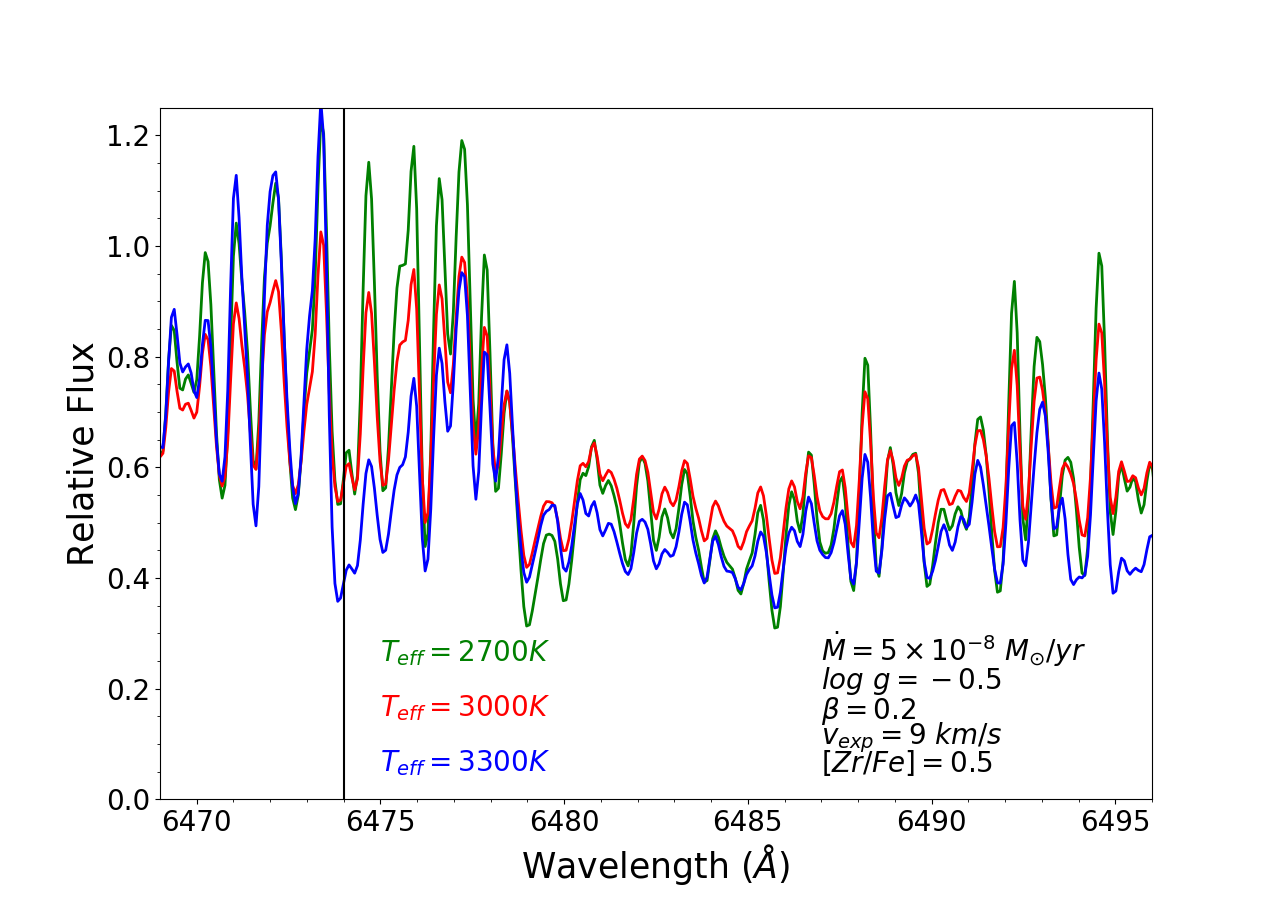}
   \includegraphics[width=9.15cm,height=6.7cm,angle=0]{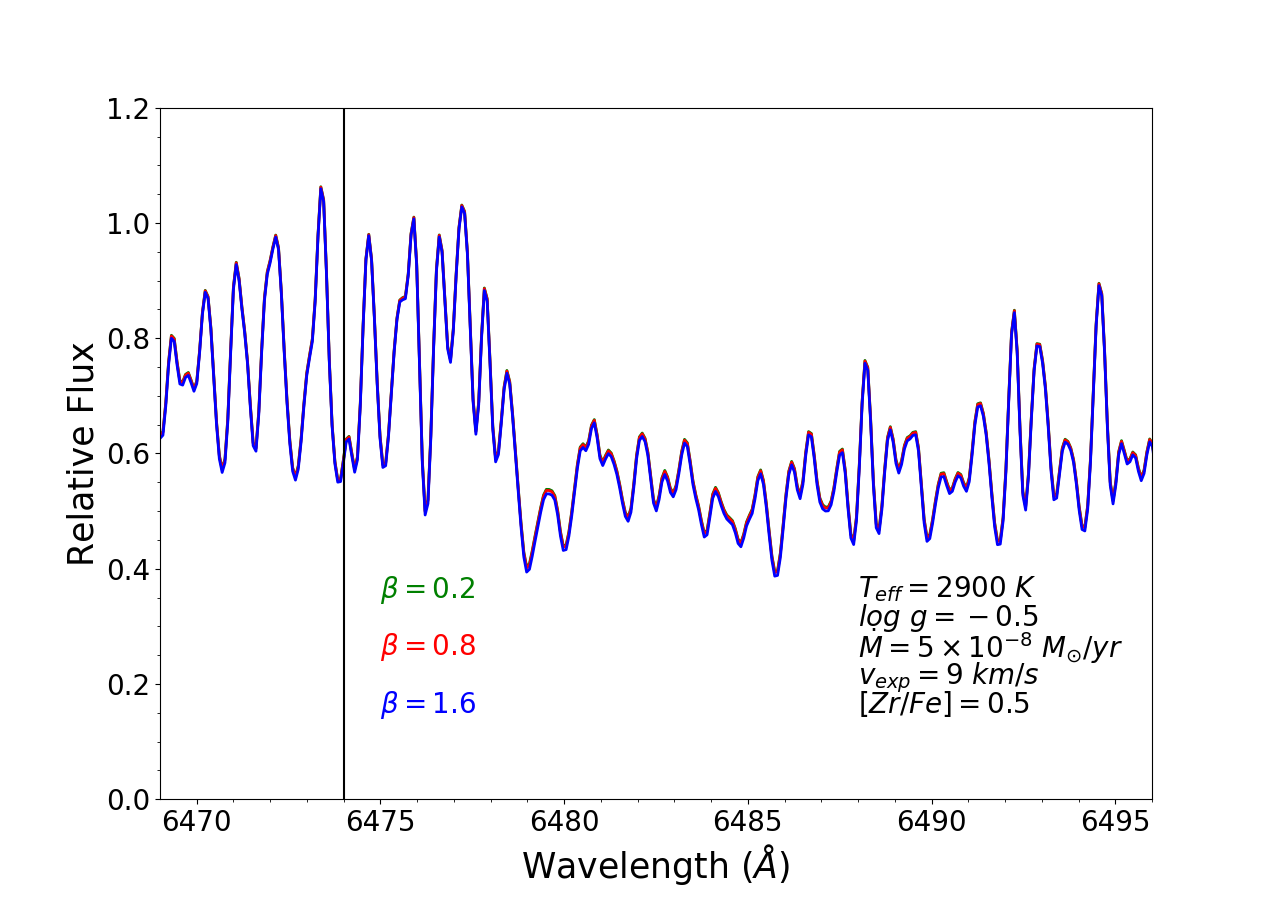}
   \includegraphics[width=9.15cm,height=6.7cm,angle=0]{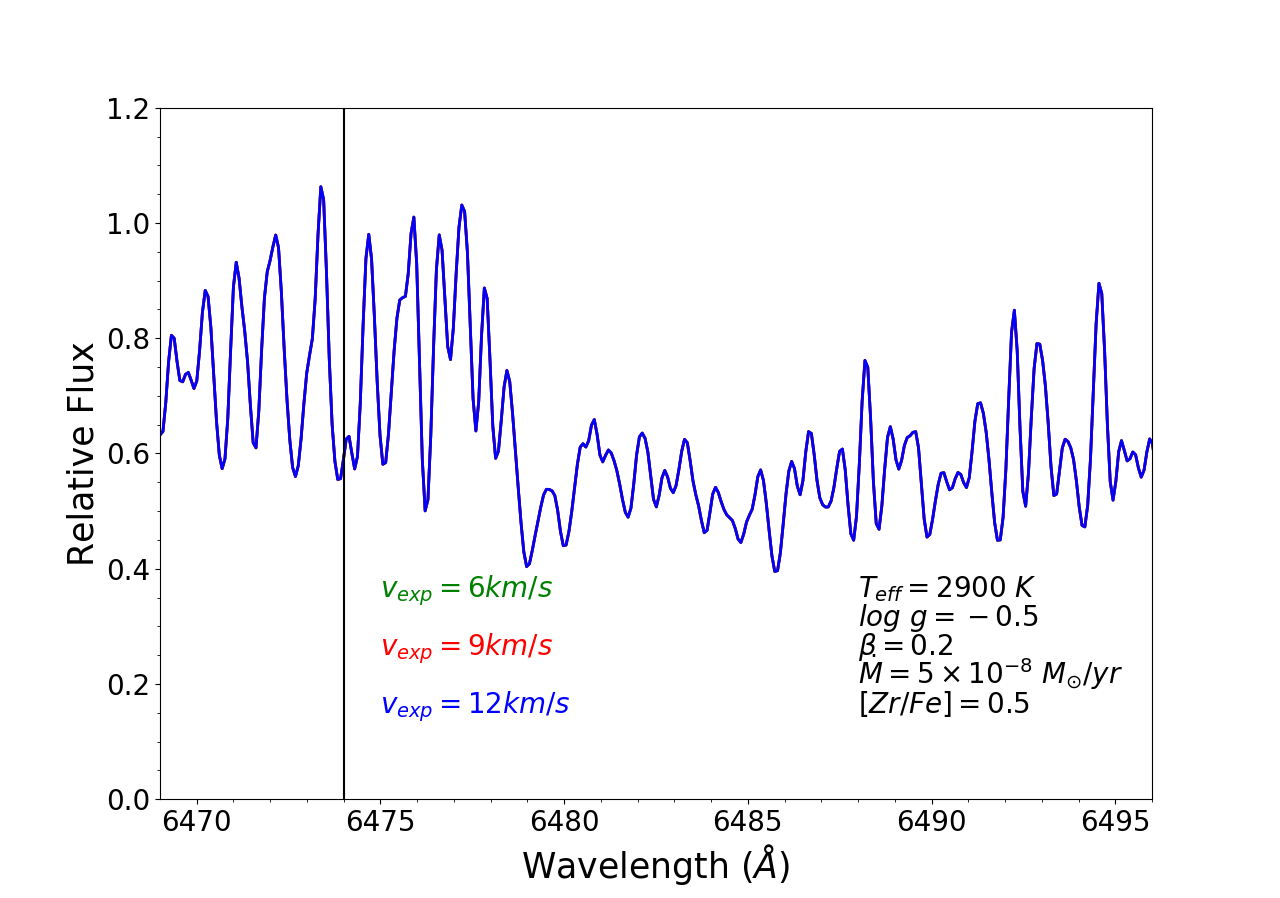}
      \caption{Illustrative examples of synthetic spectra for different stellar
(T$_{eff}$) and wind ($\dot{M}$, $\beta$, and $v_{exp}$(OH)) parameters in the
spectral region around the 6474 $\AA$ ZrO bandhead. The black vertical line
indicates the position of the 6474 $\AA$ ZrO bandhead.}
         \label{comparisons_Zr}
   \end{figure*}

Here, we analyze how the variations in stellar (T$_{eff}$) and wind
($\dot{M}$, $\beta$ and $v_{exp}$(OH)) parameters influence the output synthetic
spectra. Figures \ref{comparisons_Rb} and \ref{comparisons_Zr} show examples of
synthetic spectra for different stellar and wind parameters in the spectral
regions around the 7800 $\AA$ Rb I line and 6474 $\AA$ ZrO bandhead,
respectively. We note that the fraction of the absortion at 7800
\AA~due to other species (e.g., TiO) is tipically around 20\%.

The Rb I line profile is very sensitive to the wind mass-loss rate $\dot{M}$
(especially for $\dot{M}$ $\geq$ 10$^{-8}$ M$_{\odot}yr^{-1}$); the Rb I line is
significantly deeper and blue-shifted with increasing $\dot{M}$ (Figure
\ref{comparisons_Rb}, top-left panel). However, the Rb I line profile is much
less sensitive to changes of the wind velocity-law ($\beta$ parameter); 
being only slightly deeper with increasing $\beta$ (Figure \ref{comparisons_Rb}, 
bottom-left panel). In addition, for $\beta$ values higher
than $\sim$1.2 (shallower velocity profiles), the Rb I line profile is
not sensitive to variations of the expansion velocity $v_{exp}$(OH) because the
velocity profiles are very similar in our extended model atmosphere (up to
$\sim$10$^{14}$ cm; see Figure 1). Variations in the expansion velocity
$v_{exp}$(OH) mainly affect the blue-shift of the Rb I line and, in addition, 
for large $v_{exp}$(OH) values the core of the Rb I line is less deep (Figure
\ref{comparisons_Rb}, bottom-right panel). Finally, the Rb I absorption line is
stronger with decreasing effective temperature T$_{eff}$ (as expected; Figure
\ref{comparisons_Rb}, top-right panel) but this time the wealth of TiO molecular
lines and the pseudo-continua are also affected. We note that all these effects
(variations in the Rb I profile in terms of depth and blue-shift) are more
evident for extreme mass-loss rates ($\dot{M}$ $\geq$ 10$^{-7}$
M$_{\odot}yr^{-1}$) and higher Rb abundances.  

On the other hand, the ZrO bandhead profile is not sensitive to the wind
parameters $\dot{M}$, $\beta$, and $v_{exp}$(OH) (see Figure \ref{comparisons_Zr};
top-left, bottom-left, and bottom-right panels, respectively). The ZrO bandhead
profile (as well as the adjacent TiO lines and pseudo-continuum) are, again as
expected, stronger with decreasing T$_{eff}$ (Figure \ref{comparisons_Zr},
top-right panel). This is because ZrO is formed deeper than Rb I in the
atmosphere, being much less affected by the circumstellar envelope and radial
wind.  


\section{Results}\label{sec:results}

As we have mentioned above, there are important degeneracies in the resulting
mini-grids of synthetic spectra for each star. 
Two synthetic spectra with the same $T_{eff}$, $log g$ and $v_{exp}$(OH), but 
different $\beta$, $\dot{M}$, and [Rb/Fe] abundances could be practically identical 
in spite of the different wind parameters. This complicates the abundance
analysis because the wind model parameters $\dot{M}$ and $\beta$ are generally 
not known for
stars in our sample. In any case, we can use some observational constraints and
previous results on a few similar OH/IR stars to limit the possible variation
range of these wind parameters (in particular for the mass-loss rates $\dot{M}$,
see below). 

By using multiple rotationally excited lines of both $^{12}$CO and $^{13}$CO,
\cite{debeck10} provide accurate mass-loss rates $\dot{M}$ for a large sample
of Galactic AGB stars. Unfortunately, only one star (IRAS 20077$-$0625) from our
present sample of  Rb-rich OH/IR massive AGB stars is included in their work and
we cannot fit this star with our pseudo-dynamical models (see below). There are seven 
massive AGB stars of OH/IR type (WX Psc, V669 Cas, NV Aur, OH 26.5$+$0.6, OH 
44.8$-$2.3, IRC $-$10529 and OH 104.9$+$2.4) previously studied in the optical by
\citet{garcia-hernandez07}. Their variability periods and mass-loss rates range
from 552 to 1620 days\footnote{The variability periods of our sample stars are
also lower, from $\sim$320 to 580 days (only two stars display periods in excess
of 580 days; see Table 1).} and from 1.8$\times$10$^{-5}$ $M_{\odot}yr^{-1}$ to
9.7$\times$10$^{-6}$ $M_{\odot}yr^{-1}$, respectively. Interestingly, all these
stars are extremely obscured in the optical, being too red or without optical
counterpart\footnote{The only exception is WX Psc as already noted by
\citet{zamora14}. This star (with a mass-loss rate of $\sim$1.8 $\times$
10$^{-5}$ $M_{\odot}yr^{-1}$, \citealt{debeck10}; \citealt{justtanont13}) has an
extremely faint optical counterpart. The S/N around 7800 \AA\ is too low for
an abundance analysis but a strong Rb I absorption line is clearly detected in
its optical spectrum.}; they likely already have entered the superwind phase.
Thus, the $\dot{M}$ values in optically obscured OH/IR AGB stars can be taken as
upper limits (i.e. $<$10$^{-6}$ $M_{\odot}yr^{-1}$) for our sample of OH/IR
massive AGB stars with optical counterparts; i.e. with useful spectra around the
7800 \AA\ Rb I line. Indeed, we generally find that lower mass-loss rates
($\sim$10$^{-7}-$10$^{-8}$ $M_{\odot}yr^{-1}$) give superior fits to the
observed Rb I line profiles. Mass-loss rates of $\sim$10$^{-6}$ $M_{\odot}yr^{-1}$ 
(or higher) give strong Rb I absorption lines for solar Rb abundances (see also 
\citealt{zamora14}) with the consequence that all stars in our
sample of OH/IR massive AGB stars would be Rb-poor. By combining the variability
periods from Table 1 and the mass-loss rates estimated from the Rb I line
profiles (mainly in the range $\sim$10$^{-7}-$10$^{-8}$ $M_{\odot}yr^{-1}$; see
Table 2) into the AGB mass-loss formula by \citet{vassiliadis93} (their Eq. (5)), we obtain
reasonable current stellar masses in the range $\sim$2.5$-$6 $M_{\odot}$. 
In Table \ref{table_masses} we show the mass-loss rates obtained from the best spectral
fits ($\dot{M}_{fit}$) and the current stellar masses by using the mass-loss
expression from \cite{vassiliadis93}.

\begin{table}[]
\centering
\caption{Mass-loss rates estimated from the best spectral fits and current stellar
masses obtained by using the \cite{vassiliadis93} mass-loss formula (their Eq. (5)).}
\label{table_masses}
\renewcommand{\arraystretch}{1.25}
\begin{tabular}{cccc}
\hline
\hline
IRAS name  & Period   & $\dot{M}$            & M$_{current}$  \\
           &   (days) & ($M_{\odot}yr^{-1}$) & ($M_{\odot}$) \\
\hline
01085$+$3022 & 560      & 1.0$\times$10$^{-7}$ & 4.6   \\
04404$-$7427 & 534      & 1.0$\times$10$^{-7}$ & 4.3   \\
05027$-$2158 & 368      & 1.0$\times$10$^{-7}$ & 2.7   \\
05098$-$6422 & 394      & 5.0$\times$10$^{-7}$ & 2.4   \\
05151$+$6312 & ...      & 1.0$\times$10$^{-8}$ & ...    \\
06300$+$6058 & 440      & 1.0$\times$10$^{-7}$ & 3.4   \\
07222$-$2005 & 1200     & ...   & ...    \\
09194$-$4518 & ...      & ...    & ...    \\
10261$-$5055 & 317      & 1.0$\times$10$^{-9}$ & 3.8   \\
14266$-$4211 & 389      & 5.0$\times$10$^{-8}$ & 3.1   \\
15193$+$3132 & 360      & 1.0$\times$10$^{-9}$ & 4.2   \\
15576$-$1212 & 415      & 1.0$\times$10$^{-8}$ & 3.9   \\
16030$-$5156 & 579      & 1.0$\times$10$^{-8}$ & 5.6   \\
16037$-$1024 & 360      & 1.0$\times$10$^{-7}$ & 2.6   \\
17034$-$1024 & 346      & 1.0$\times$10$^{-8}$ & 3.2   \\
18429$-$1721 & 481      & 1.0$\times$10$^{-8}$ & 4.6   \\
19059$-$2219 & 510      & 5.0$\times$10$^{-8}$ & 4.3   \\
19426$+$4342 & ...      & ...    & ...    \\
20052$+$0554 & 450      & 5.0$\times$10$^{-7}$ & 2.9   \\
20077$-$0625 & 680      & ...    & ...    \\
20343$-$3020 & 349      & 1.0$\times$10$^{-9}$ & 4.1   \\
\hline
\end{tabular}
\end{table}

The $\beta$ parameter in our models (only up to $\sim$10$^{14}$ cm from the
photosphere; see Figure 1) cannot be directly compared with other estimations of
this parameter in the literature \citep[e.g.][]{decin10, Danilovich15}, which
map much outer regions in the circumstellar envelope and that usually obtain 
quite high and uncertain values (0 $\leq$ $\beta$ $\leq$ 5.0). However, the 
effect of the $\beta$ parameter on our synthetic spectra is minor compared 
to the mass-loss rate $\dot{M}$ and we keep it as a free parameter in our 
abundance analysis. We note also that the velocity profiles are very similar
in our extended model atmosphere for $\beta$ $\geq$ 1.2; i.e., the Rb I line
profile is not more sensitive to variations of the expansion velocity and the
abundance results are very similar for $\beta$ $\geq$ 1.2. We generally
find better fits with low $\beta$ values (or steeper velocity profiles; see
Table \ref{table_beta_free}). 

As mentioned above, the parameters of the hydrostatic models providing
the best fit to the observations and the Rb abundances derived are shown in
Table \ref{table_obs_param}.  The static models use the solar abundances from
\cite{grevesse98} for computing the Rb abundances
\citep[see][]{garcia-hernandez06, garcia-hernandez07}, while our pseudo-dynamical
models use the more recent solar abundances from \cite{grevesse07}. In
\cite{zamora14} we compared the Rb abundances from static models using
\cite{grevesse98} and \cite{grevesse07}, and the Rb abundances obtained agree
within $\sim$0.2 dex in most cases. 

\begin{figure*}
\centering
\includegraphics[width=9.1cm,height=6.7cm,angle=0]{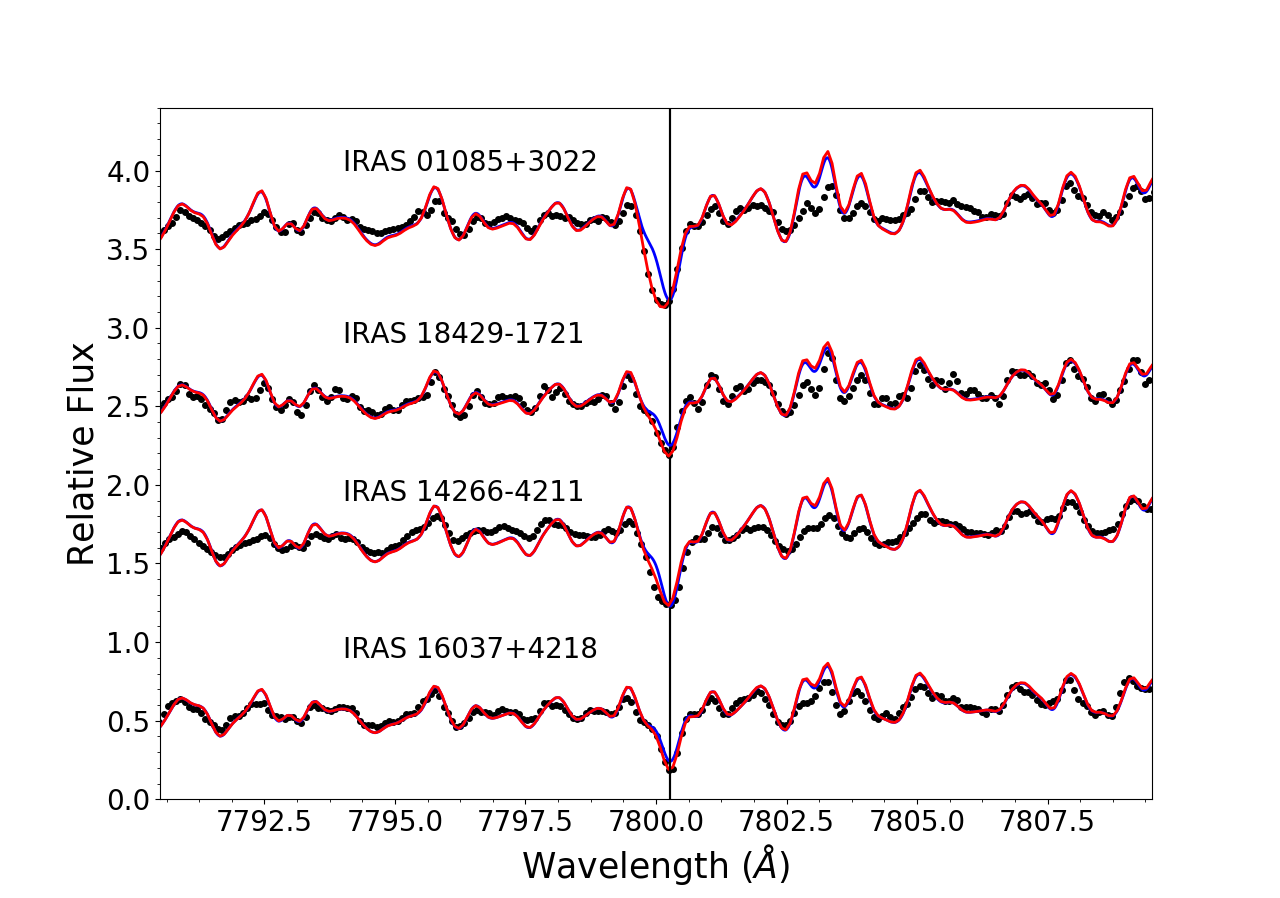}
\includegraphics[width=9.1cm,height=6.7cm,angle=0]{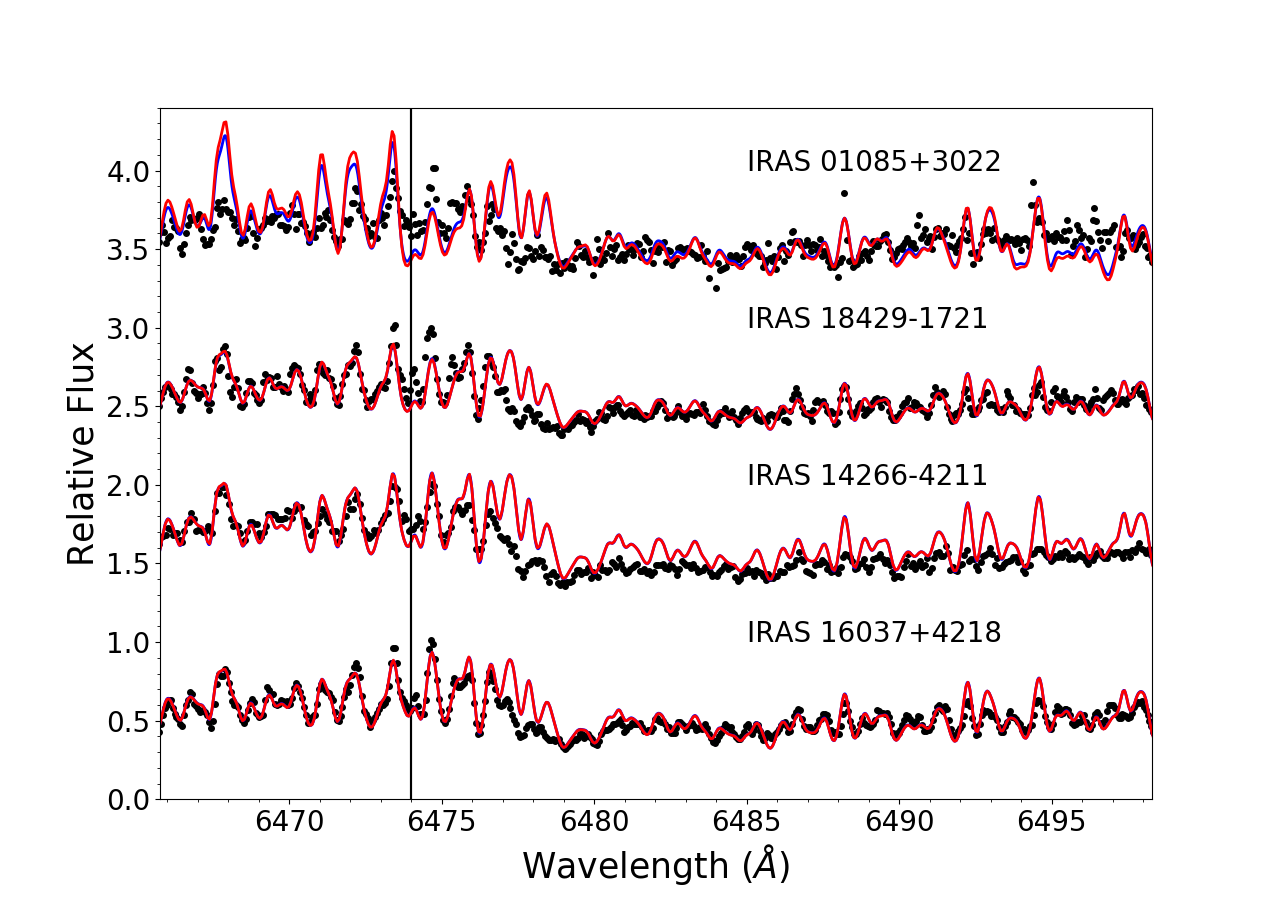} 
\caption{The Rb I 7800 $\AA$ (\textit{left panel}) and ZrO 6474 $\AA$
(\textit{right panel}) spectral regions in massive Galactic AGB stars. The
pseudo-dynamical models (red lines) that best fit the observations (black dots) are
shown in four sample stars. For comparison, the hydrostatic models are also
displayed (blue lines). The location of the Rb I line and the ZrO bandhead are
indicated by black vertical lines.}
\label{Rb_Zr}
\end{figure*}

\begin{figure*}
\centering
\includegraphics[width=9.15cm]{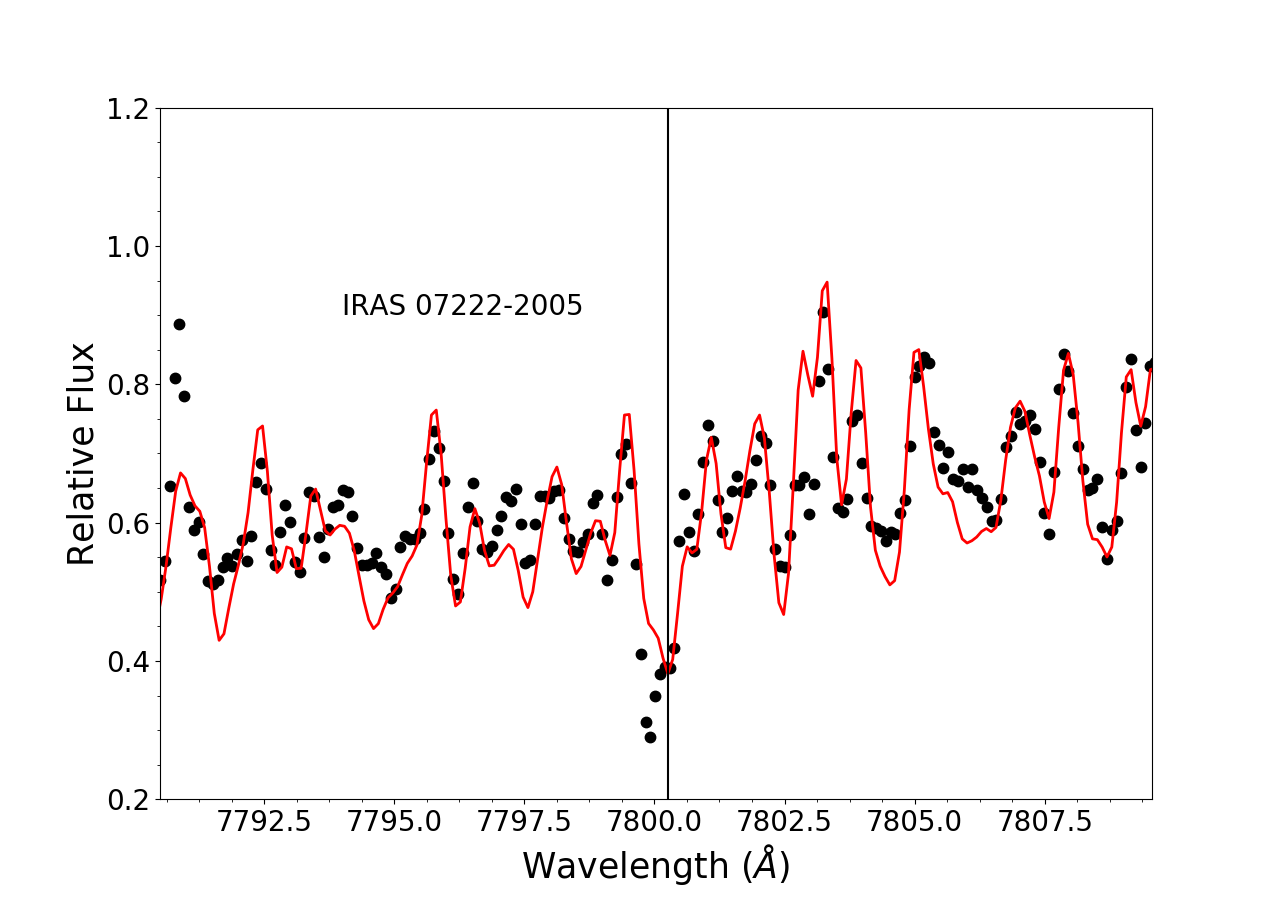}
\includegraphics[width=9.15cm]{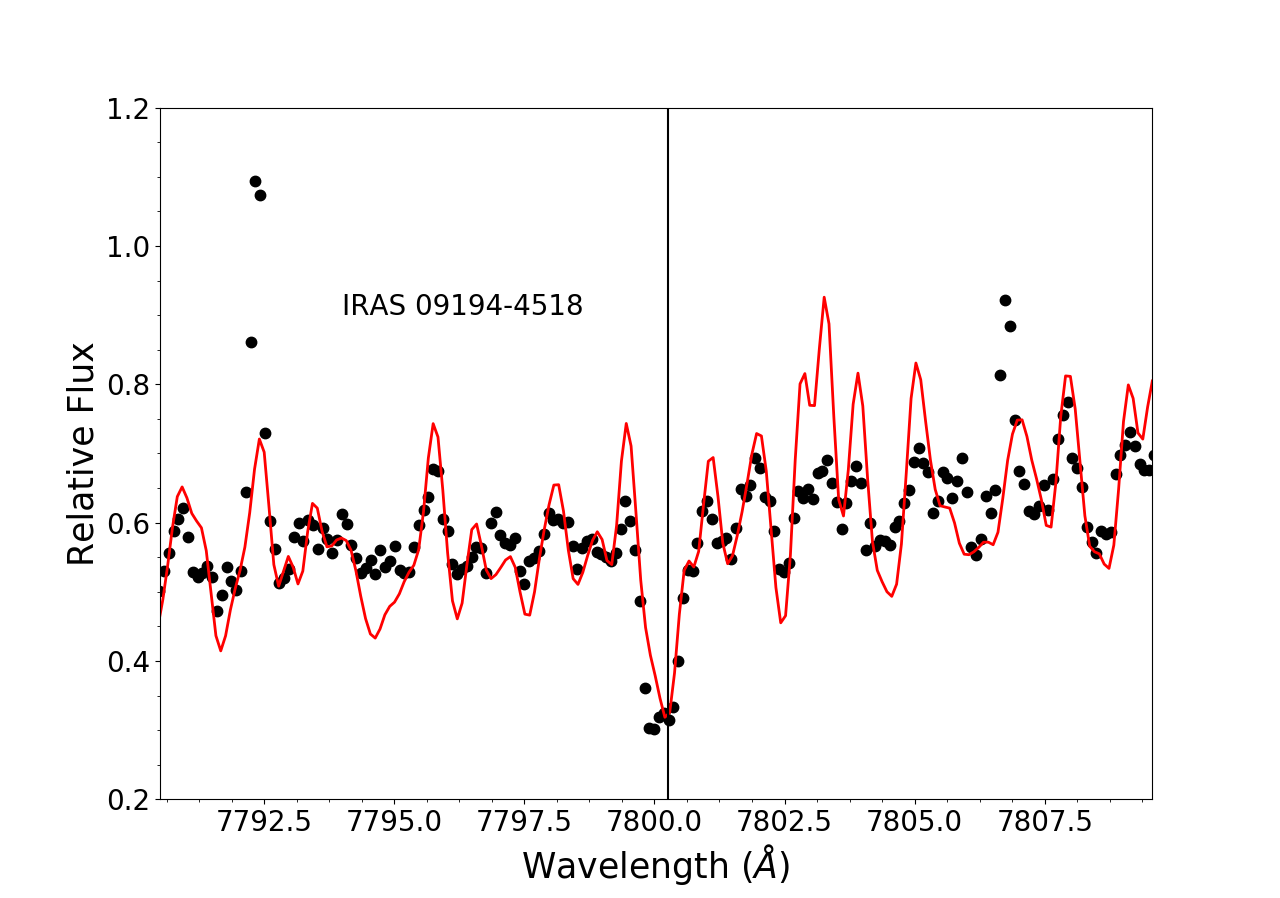}
\caption{IRAS 07222$-$2005 (\textit{left panel}) and IRAS 09194$-$4518
(\textit{right panel}) display two components (circumstellar and photospheric) 
in the 7800 $\AA$ Rb I line (black line) and cannot be reproduced by our present 
version of the spectral synthesis code. The red line shows the pseudo-dynamical 
synthetic spectrum that gives a good fit to the photospheric component only.}
\label{no_ajustado}
\end{figure*}

Figure \ref{Rb_Zr} shows that our pseudo-dynamical atmosphere models reproduce the
observed 7800 $\AA$ Rb I line profile much better than the classical hydrostatic
models in four sample stars (see Appendix \ref{Append_sample} for the rest of sample stars). 
On the other hand, the Zr abundances derived from the extended models
are similar to those obtained with the hydrostatic models because the 6474 $\AA$
ZrO bandhead is formed deeper in the atmosphere and is less affected by the
radial velocity field \citep{zamora14}. We could obtain the Rb and Zr abundances
(or upper limits) for 17 sample stars. The rest of sample stars (IRAS
07222$-$2005, IRAS 09194$-$4518, IRAS 19426$+$4342 and IRAS 20077$-$0625) seem
to display different Rb I line profiles (e.g., with more than one circumstellar
contribution or anomalously broad profiles with red-extended wings; see e.g.,
Figure \ref{no_ajustado}) that cannot be completely reproduced by our present
version of the spectral synthesis code. In the two stars (IRAS 07222$-$2005 and 
IRAS 09194$-$4518) shown in Figure \ref{no_ajustado} we cannot fit the two Rb I
components (circumstellar and photospheric) at the same time; e.g., we only might
fit partially the blue-shifted circumstellar component using a larger
mass-loss rates  ($>$10$^{-6}$ $M_{\odot}yr^{-1}$). Curiously, these two stars
present the largest  periods (see Table \ref{table_obs_param}) and they may be
the more extreme and evolved stars in our sample, where our extended models do
not not work so well (e.g., due to even more extended atmospheres). It is not 
completely clear,
however, if the observed profiles are real because these four sample stars 
have the lowest quality spectra (\textit{S/N} $<$ 30 at 7800 \AA; see Table 
\ref{table_obs_param}). 

For the two sample stars with unknown OH expansion velocity, IRAS 16030$-$5156 
and IRAS 17034$-$1024, we explore the velocity range displayed by other stars 
with similar variability periods (see Table 
\ref{table_obs_param}).  Similar fits can be obtained for $v_{exp}$(OH) 
$\sim$ 7$-$12 and 7$-$9 kms$^{-1}$ (in combination with sligthly different 
wind parameters) for IRAS 16030$-$5156 and IRAS 17034$-$1024, respectively, 
and we thus adopt average velocities of 10 and 8 kms$^{-1}$, respectively, in 
the abundance analysis (see Table \ref{table_beta_free}).

\begin{figure}
   \centering
   \includegraphics[width=9.55cm,angle=0]{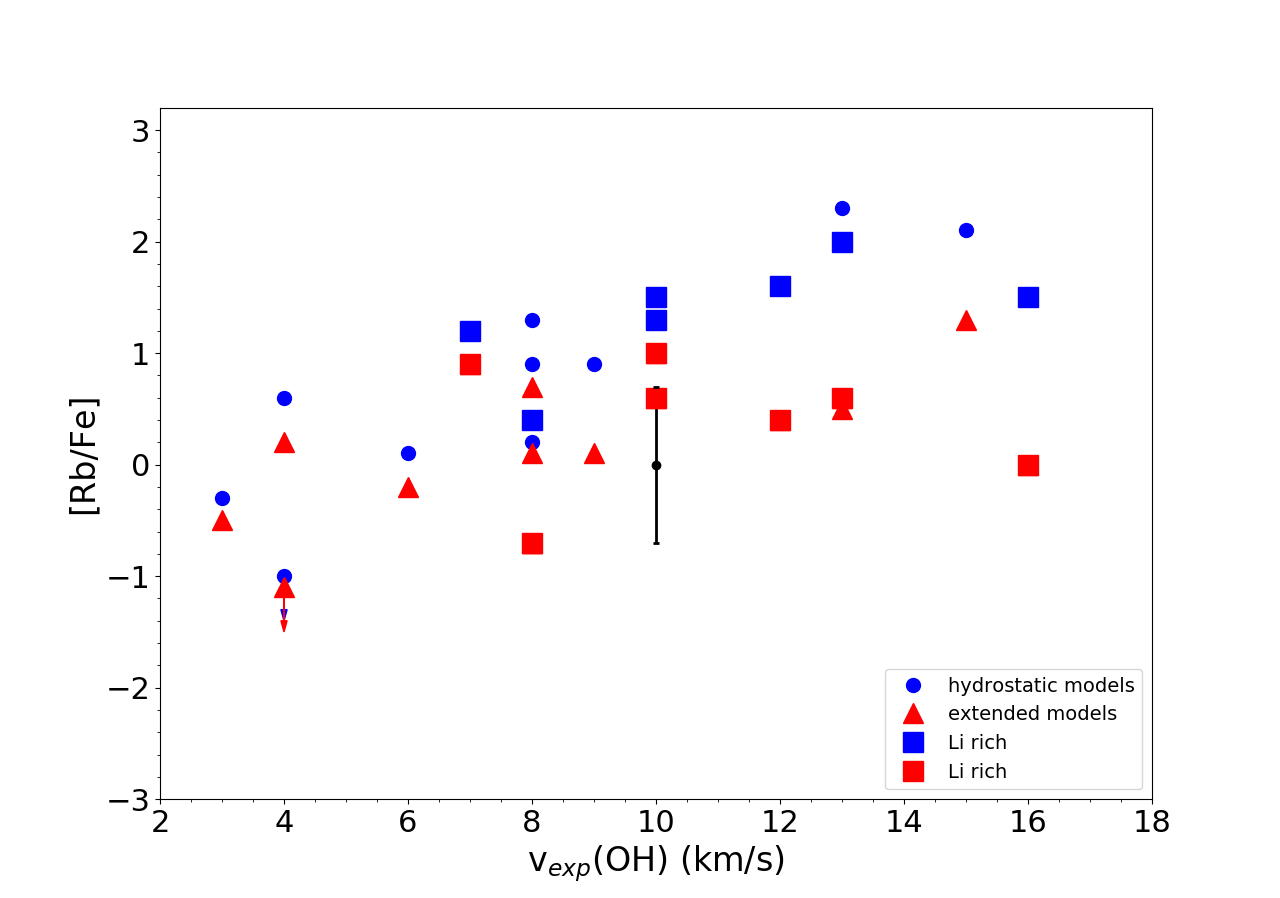}
      		\caption{Rb abundances derived both with hydrostatic (blue dots) 
      		and pseudo-dynamical model atmospheres (red triangles) with best-fit 
      		parameters plotted against the OH expansion velocity. The Li-rich stars 
      		are indicated by squares. A typical error bar of $\pm$0.7 dex is also displayed.}
         \label{Rb_beta_free}
\end{figure}

Table \ref{table_beta_free} shows the atmospheric and wind parameters as well as
the Rb and Zr abundances (or upper limits) from the best fits to the observed
spectra when the wind parameters $\dot{M}$ and $\beta$ are not fixed. In most
cases, the best fit is obtained for both low $\beta$ ($\sim$0.2) and $\dot{M}$
($\sim$10$^{-9}$-10$^{-7}$ M$_{\odot}$ yr$^{-1}$) values. The new Rb abundances
obtained from extended models are lower than those obtained using the hydrostatic 
models, and the difference is larger for stars with higher hydrostatic Rb
abundances. In addition, this difference is smaller for lower $v_{exp}$(OH) and
increases with increasing $v_{exp}$(OH), as expected. On the other hand, in the
case of Zr we obtain upper limits mostly between 0.0 and $+$0.25 dex, as derived 
from the hydrostatic models. Figure \ref{Rb_beta_free} displays the hydrostatic 
and pseudo-dynamical Rb abundances versus the OH expansion velocity  for the wind 
parameters that provide the best fits (Table \ref{table_beta_free}. 
We plot the Rb abundances versus the expansion 
velocity because the $v_{exp}$(OH) can be used as a mass indicator independent 
of the distance in OH/IR stars \citep{garcia-hernandez07}. 
In addition, in Figure \ref{Rb_beta_free} we have marked the Li-rich stars 
\citep{garcia-hernandez07} by squares. About half of the stars with v$_{exp}$(OH) 
$>$ 6 km/s are Li-rich and most of these stars are also the more Rb-rich ones.
We get pseudo-dynamical abundances lower than the hydrostatic ones and a
worse correlation between\footnote{The correlation coefficients are $r=$ 0.84 and 
0.54 for the hydrostatic and pseudo-dynamic cases, respectively.} the Rb abundances 
and $v_{exp}$(OH); the Rb-$v_{exp}$(OH) relationship is flatter (with a higher
degeneration) for the pseudo-dynamical case (see also Sect. \ref{comparison_section}).

\begin{table*}
\centering
\caption{Atmosphere parameters and best-fit Rb pseudo-dynamical abundances for 
the listed wind parameters $\dot{M}$ and $\beta$. The asterisks indicate that 
the best fitting $T_{eff}$ in the ZrO 6474 $\AA$ spectral region is warmer 
(3300 K) than the one around the Rb I 7800 $\AA$ line \citep{garcia-hernandez06, 
garcia-hernandez07}.}\label{table_beta_free} 
\renewcommand{\arraystretch}{1.25}
\begin{tabular}{ccccccccc}
\hline
\hline
IRAS name  & $T_{eff}$ (K) & log $g$ & $\beta$ & $\dot{M}$ (M$_{\odot}$ yr$^{-1}$) 
& $v_{exp}$(OH) (km s$^{-1}$) & {[}Rb/Fe{]}$_{static}$ & {[}Rb/Fe{]}$_{dyn}$ & {[}Zr/Fe{]}$_{dyn}$ \\
\hline
01085$+$3022 & 3000* & $-$0.5 & 0.2 & 1.0$\times$10$^{-7}$ & 13 & 2.0  & 0.6 & $\leq$ 0.0 \\
04404$-$7427 & 3000  & $-$0.5 & 0.2 & 1.0$\times$10$^{-7}$ & 8  & 1.3  & 0.1 & $\leq$ 0.0 \\
05027$-$2158 & 2800	 & $-$0.5 & 0.4 & 1.0$\times$10$^{-7}$ & 8  & 0.4  & $-$0.7 & $\leq$ +0.5 \\
05098$-$6422 & 3000  & $-$0.5 & 1.4 & 1.0$\times$10$^{-8}$ & 6  & 0.1  & $-$0.2 & $\leq$ +0.25 \\
05151$+$6312 & 3000  & $-$0.5 & 1.0 & 1.0$\times$10$^{-8}$ & 15 & 2.1  & 1.3 & $\leq$ +0.25 \\
06300$+$6058 & 3000  & $-$0.5 & 0.2 & 1.0$\times$10$^{-7}$ & 12 & 1.6  & 0.4 & $\leq$ 0.0 \\
07222$-$2005 & 3000  & $-$0.5 & ... & ... & 8  & 0.6  & ... & ... \\
09194$-$4518 & 3000  & $-$0.5 & ... & ... & 11 & 1.1  & ... & ... \\
10261$-$5055 & 3000  & $-$0.5 & 0.2 & 1.0$\times$10$^{-9}$ & 4 & $<$ $-$1.0 & $<$ $-$1.1 & $\leq$ +0.25 \\
14266$-$4211 & 2900  & $-$0.5 & 0.2 & 5.0$\times$10$^{-8}$ & 9  & 0.9  & 0.1 & $\leq$ 0.0 \\
15193$+$3132 & 2800  & $-$0.5 & 1.6 & 1.0$\times$10$^{-9}$ & 3  & $-$0.3  & $-$0.5 & $\leq$ 0.0 \\
15576$-$1212 & 3000  & $-$0.5 & 0.2 & 1.0$\times$10$^{-8}$ & 10 & 1.5  & 1.0 & $\leq$ 0.0 \\
16030$-$5156 & 3000  & $-$0.5 & 0.2 & 1.0$\times$10$^{-8}$ & 10 & 1.3  & 0.6 & $\leq$ +0.25 \\
16037$+$4218 & 2900  & $-$0.5 & 1.2 & 1.0$\times$10$^{-8}$ & 4  & 0.6  & 0.2 & $\leq$ +0.25\\
17034$-$1024 & 3300  & $-$0.5 & 0.8 & 1.0$\times$10$^{-8}$ & 8 & 0.2  & $-$0.7 & $\leq$ 0.0 \\
18429$-$1721 & 3000  & $-$0.5 & 0.2 & 1.0$\times$10$^{-8}$ & 7  & 1.2  & 0.9 & $\leq$ +0.25 \\
19059$-$2219 & 3000  & $-$0.5 & 0.2 & 1.0$\times$10$^{-7}$ & 13 & 2.3  & 0.5 & $\leq$ +0.25 \\
19426$+$4342 & 3000  & $-$0.5 & ... & ... & 9  & 1.0  & ... & ... \\
20052$+$0554 & 3000* & $-$0.5 & 0.2 & 5.0$\times$10$^{-7}$ & 16 & 1.5  & 0.0 & $\leq$ 0.0 \\
20077$-$0625 & 3000  & $-$0.5 & ... & ... & 12 & 1.3  & ... & ... \\
20343$-$3020 & 3000  & $-$0.5 & 1.2 & 1.0$\times$10$^{-9}$ & 8  & 0.9  & 0.7 & $\leq$ 0.0 \\
\hline  
\end{tabular}
\tablefoot{We have checked the sensitivity of the derived abundances to small changes in 
the model atmosphere parameters ($\Delta T_{eff}$ = $\pm100$ K, $\Delta \beta$ = 0.2, 
$\Delta log (\dot{M}/M_{\odot}$yr$^{-1})$ = 0.5, $\Delta v_{exp}$(OH) = 5 km s$^{-1}$). 
The dominant sources of uncertainties are $\dot{M}$ and $T_{eff}$ for Rb and Zr, 
respectively, which result in error bars of $\pm$0.7 and $\pm$0.3 dex, respectively. 
In the hydrostatic case, the formal Rb uncertainties (quoted by \cite{garcia-hernandez06}) 
due to changes of all the atmosphere parameters are $\pm$0.8 dex.}
\end{table*}

\begin{table*}
\centering
\caption{Atmosphere parameters and Rb pseudo-dynamical abundances with $\beta$ = 
0.2 and $\beta$ = 1.2, respectively. The asterisks indicate that the best fitting 
$T_{eff}$ in the ZrO 6474 $\AA$ spectral region is warmer (3300 K) than the one 
around the Rb I 7800 $\AA$ line \citep{garcia-hernandez06, garcia-hernandez07}.}
\label{table_beta_test}
\renewcommand{\arraystretch}{1.25}
\begin{tabular}{ccccccccc}
\hline
\hline
IRAS name  & $T_{eff}$ (K) & log $g$ & $\beta$ & $\dot{M}$ (M$_{\odot}$ yr$^{-1}$) 
& $v_{exp}$(OH) (km s$^{-1}$) & {[}Rb/Fe{]}$_{static}$ & {[}Rb/Fe{]}$_{dyn}$ & 
{[}Zr/Fe{]}$_{dyn}$\\
\hline
01085$+$3022 & 3000* & $-$0.5 & \begin{tabular}[c]{@{}c@{}}0.2\\ 1.2\end{tabular} 
& \begin{tabular}[c]{@{}c@{}}1.0$\times$10$^{-7}$\\ 1.0$\times$10$^{-7}$\end{tabular} 
& 13 & 2.0  & \begin{tabular}[c]{@{}c@{}}0.6 \\ 0.3 \end{tabular} & $\leq$ 0.0 \\
\hline

04404$-$7427 & 3000  & $-$0.5 & \begin{tabular}[c]{@{}c@{}}0.2\\ 1.2\end{tabular} 
& \begin{tabular}[c]{@{}c@{}}1.0$\times$10$^{-7}$\\ 1.0$\times$10$^{-7}$\end{tabular} 
& 8  & 1.3  & \begin{tabular}[c]{@{}c@{}}0.1 \\ 0.1 \end{tabular} & 
\begin{tabular}[c]{@{}c@{}}$\leq$ 0.0 \\ \end{tabular} \\
\hline

05027$-$2158 & 2800	 & $-$0.5 & \begin{tabular}[c]{@{}c@{}}0.2\\ 1.2\end{tabular} 
& \begin{tabular}[c]{@{}c@{}}1.0$\times$10$^{-7}$\\ 1.0$\times$10$^{-7}$\end{tabular} 
& 8  & 0.4  & \begin{tabular}[c]{@{}c@{}}$-$0.6 \\ $-$1.1 \end{tabular} & 
\begin{tabular}[c]{@{}c@{}}$\leq$ +0.5 \\ \end{tabular}\\
\hline

05098$-$6422 & 3000  & $-$0.5 & \begin{tabular}[c]{@{}c@{}}0.2\\ 1.2\end{tabular} 
& \begin{tabular}[c]{@{}c@{}}1.0$\times$10$^{-8}$\\ 5.0$\times$10$^{-7}$\end{tabular} 
& 6  & 0.1  & \begin{tabular}[c]{@{}c@{}}$-$0.2 \\ $-$2.1 \end{tabular} & 
\begin{tabular}[c]{@{}c@{}}$\leq$ +0.25 \\  \end{tabular}\\
\hline

05151$+$6312 & 3000  & $-$0.5 & \begin{tabular}[c]{@{}c@{}}0.2\\ 1.2\end{tabular} 
& \begin{tabular}[c]{@{}c@{}}5.0$\times$10$^{-7}$\\ 5.0$\times$10$^{-7}$\end{tabular} 
& 15 & 2.1  & \begin{tabular}[c]{@{}c@{}}0.0 \\ $-$0.5 \end{tabular} & 
\begin{tabular}[c]{@{}c@{}}$\leq$ +0.25 \\  \end{tabular}\\
\hline

06300$+$6058 & 3000  & $-$0.5 & \begin{tabular}[c]{@{}c@{}}0.2\\ 1.2\end{tabular} 
& \begin{tabular}[c]{@{}c@{}}1.0$\times$10$^{-7}$\\ 1.0$\times$10$^{-7}$\end{tabular} 
& 12 & 1.6  & \begin{tabular}[c]{@{}c@{}}0.4 \\ 0.3 \end{tabular} & 
\begin{tabular}[c]{@{}c@{}}$\leq$ 0.0 \\  \end{tabular}\\
\hline

07222$-$2005 & 3000  & $-$0.5 & ... & ... & 8  & 0.6  & ... & ...  \\
\hline

09194$-$4518 & 3000  & $-$0.5 & ... & ... & 11 & 1.1  & ... & ... \\
\hline

10261$-$5055 & 3000  & $-$0.5 & \begin{tabular}[c]{@{}c@{}}0.2\\ 1.2\end{tabular} 
& \begin{tabular}[c]{@{}c@{}}1.0$\times$10$^{-9}$\\ 1.0$\times$10$^{-9}$\end{tabular} 
& 4  & $<$ $-$1.0 & \begin{tabular}[c]{@{}c@{}}$<$ $-$1.1 \\ $<$ $-$1.1 \end{tabular} 
& \begin{tabular}[c]{@{}c@{}}$\leq$ +0.25 \\  \end{tabular}\\
\hline

14266$-$4211 & 2900  & $-$0.5 & \begin{tabular}[c]{@{}c@{}}0.2\\ 1.2\end{tabular} 
& \begin{tabular}[c]{@{}c@{}}5.0$\times$10$^{-8}$\\ 1.0$\times$10$^{-7}$\end{tabular} 
& 9  & 0.9  & \begin{tabular}[c]{@{}c@{}}0.1  \\ $-$0.4 \end{tabular} & 
\begin{tabular}[c]{@{}c@{}}$\leq$ 0.0 \\$\leq$ $-$0.25 \end{tabular} \\
\hline

15193$+$3132 & 2800  & $-$0.5 & \begin{tabular}[c]{@{}c@{}}0.2\\ 1.2\end{tabular} 
& \begin{tabular}[c]{@{}c@{}}1.0$\times$10$^{-9}$\\ 1.0$\times$10$^{-9}$\end{tabular} 
& 3  & $-$0.3  & \begin{tabular}[c]{@{}c@{}}$-$0.4 \\ $-$0.4 \end{tabular} & $\leq$ 0.0 \\ 
\hline

15576$-$1212 & 3000  & $-$0.5 & \begin{tabular}[c]{@{}c@{}}0.2\\ 1.2\end{tabular} 
& \begin{tabular}[c]{@{}c@{}}1.0$\times$10$^{-8}$\\ 1.0$\times$10$^{-8}$\end{tabular} 
& 10 & 1.5  & \begin{tabular}[c]{@{}c@{}}1.0 \\ 0.9 \end{tabular} & 
\begin{tabular}[c]{@{}c@{}}$\leq$ 0.0 \\  \end{tabular} \\
\hline

16030$-$5156 & 3000  & $-$0.5 & \begin{tabular}[c]{@{}c@{}}0.2\\ 1.2\end{tabular} 
& \begin{tabular}[c]{@{}c@{}}1.0$\times$10$^{-8}$\\ 1.0$\times$10$^{-7}$\end{tabular} 
& \begin{tabular}[c]{@{}c@{}}10\\ \end{tabular} & 1.3  & \begin{tabular}[c]{@{}c@{}}0.6 
\\ $-$0.4 \end{tabular}  & \begin{tabular}[c]{@{}c@{}}$\leq$ +0.25 \\  \end{tabular} \\
\hline

16037$+$4218 & 2900  & $-$0.5 & \begin{tabular}[c]{@{}c@{}}0.2\\ 1.2\end{tabular} 
& \begin{tabular}[c]{@{}c@{}}1.0$\times$10$^{-8}$\\ 1.0$\times$10$^{-8}$\end{tabular} 
& 4  & 0.6  & \begin{tabular}[c]{@{}c@{}}0.5 \\ 0.2 \end{tabular} & 
\begin{tabular}[c]{@{}c@{}}$\leq$ +0.25 \\ \end{tabular} \\
\hline

17034$-$1024 & 3300  & $-$0.5 & \begin{tabular}[c]{@{}c@{}}0.2\\ 1.2\end{tabular} 
& \begin{tabular}[c]{@{}c@{}}1.0$\times$10$^{-8}$\\ 1.0$\times$10$^{-8}$\end{tabular} 
& \begin{tabular}[c]{@{}c@{}}8\\ \end{tabular} & 0.2  & \begin{tabular}[c]{@{}c@{}}$-$0.7 
\\ $-$0.8 \end{tabular} & \begin{tabular}[c]{@{}c@{}}$\leq$ 0.0 \\ \end{tabular}\\
\hline

18429$-$1721 & 3000  & $-$0.5 & \begin{tabular}[c]{@{}c@{}}0.2\\ 1.2\end{tabular} 
& \begin{tabular}[c]{@{}c@{}}1.0$\times$10$^{-8}$\\ 5.0$\times$10$^{-7}$\end{tabular} 
& 7  & 1.2  & \begin{tabular}[c]{@{}c@{}}0.9 \\ $-$1.0 \end{tabular} & 
\begin{tabular}[c]{@{}c@{}}$\leq$ +0.25 \\ $\leq$ 0.0 \end{tabular} \\
\hline

19059$-$2219 & 3000  & $-$0.5 & \begin{tabular}[c]{@{}c@{}}0.2\\ 1.2\end{tabular} 
& \begin{tabular}[c]{@{}c@{}}1.0$\times$10$^{-7}$\\ 5.0$\times$10$^{-8}$\end{tabular} 
& 13 & 2.3 & \begin{tabular}[c]{@{}c@{}}0.5 \\ 0.5 \end{tabular} & 
\begin{tabular}[c]{@{}c@{}}$\leq$ +0.25 \\ \end{tabular} \\
\hline

19426$+$4342 & 3000  & $-$0.5 & ... & ... & 9  & 1.0  & ... & ... \\
\hline

20052$+$0554 & 3000* & $-$0.5 & \begin{tabular}[c]{@{}c@{}}0.2\\ 1.2\end{tabular} 
& \begin{tabular}[c]{@{}c@{}}5.0$\times$10$^{-7}$\\ 5.0$\times$10$^{-7}$\end{tabular} 
& 16 & 1.5  & \begin{tabular}[c]{@{}c@{}}0.0 \\ $-$0.7 \end{tabular}  & 
\begin{tabular}[c]{@{}c@{}}$\leq$ 0.0 \\ \end{tabular} \\
\hline

20077$-$0625 & 3000  & $-$0.5 & ... & ... & 12 & 1.3  & ... & ... \\
\hline

20343$-$3020 & 3000  & $-$0.5 & \begin{tabular}[c]{@{}c@{}}0.2\\ 1.2\end{tabular} 
& \begin{tabular}[c]{@{}c@{}}1.0$\times$10$^{-9}$\\ 1.0$\times$10$^{-9}$\end{tabular} 
& 8  & 0.9  & \begin{tabular}[c]{@{}c@{}}0.7 \\ 0.7 \end{tabular} & 
\begin{tabular}[c]{@{}c@{}}$\leq$ 0.0 \\ \end{tabular}  \\
\hline  
\end{tabular}
\end{table*}

\begin{figure}
   \centering
   \includegraphics[width=9.55cm,angle=0]{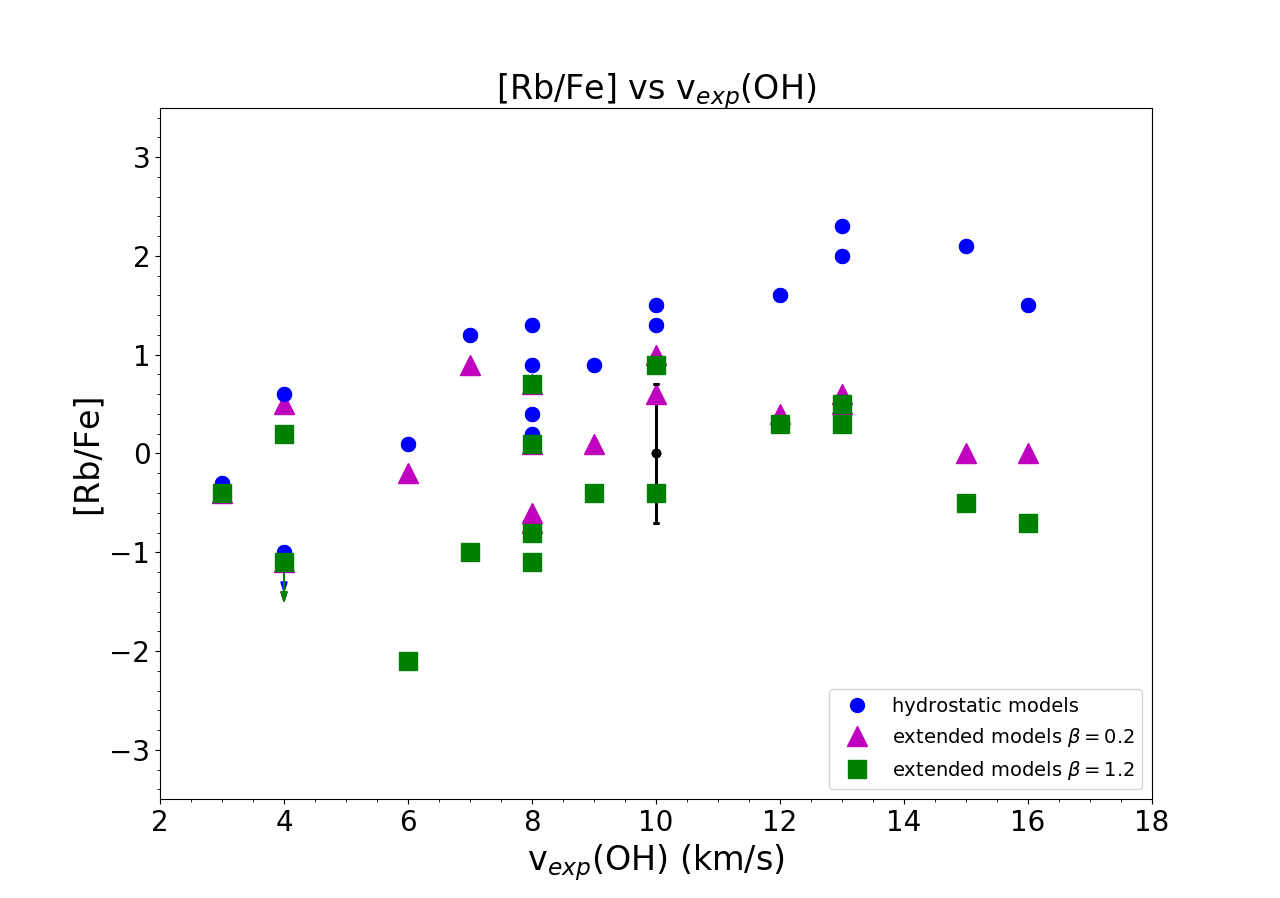}
\caption{Rb abundances derived both with hydrostatic (blue dots) and
extended model atmospheres with $\beta$ = 0.2 (magenta triangles) and
$\beta$ = 1.2 (green squares) plotted against the OH expansion
velocity.}
         \label{Rb_beta_fixed}
\end{figure}

Also, we have carried out several tests with different $\beta$ and $\dot{M}$
values in order to check the sensitivity of the derived abundances to variations
of the wind parameters. In Table \ref{table_beta_test} we present the wind
parameters and Rb abundances obtained when fixing $\beta$ to 0.2 and 1.2.
Basically, the Rb abundances are lower in the $\beta$ = 1.2 case because a
higher $\beta$ deepens the Rb I 7800 $\AA$ line for the same Rb abundance (see 
Figure \ref{comparisons_Rb}); in a few cases, however, the $\dot{M}$ of the best
fit also changes, further affecting the determination of the Rb abundance. The 
Zr abundances (or upper limits) are similar in most cases; the upper limits only change 
when the $\dot{M}$ is not the same for the $\beta$ = 0.2 and 1.2 cases. Figure
\ref{Rb_beta_fixed} represents the Rb abundances obtained versus $v_{exp}$(OH) 
for $\beta$ = 0.2 
and $\beta$ = 1.2
. By comparing the Rb abundances from Figure
\ref{Rb_beta_free} and Figure \ref{Rb_beta_fixed}, it is clear that the Rb
abundances are slightly lower in the $\beta$ = 1.2 case. Moreover, the correlation
between the pseudo-dynamical Rb abundances and $v_{exp}$(OH) for different $\beta$ values
is worse (e.g., flatter) than the hydrostatic case. In addition, the dispersion seems 
to be larger for the $\beta$ = 1.2 case.

\begin{figure}
   \centering
   \includegraphics[width=9.55cm,angle=0]{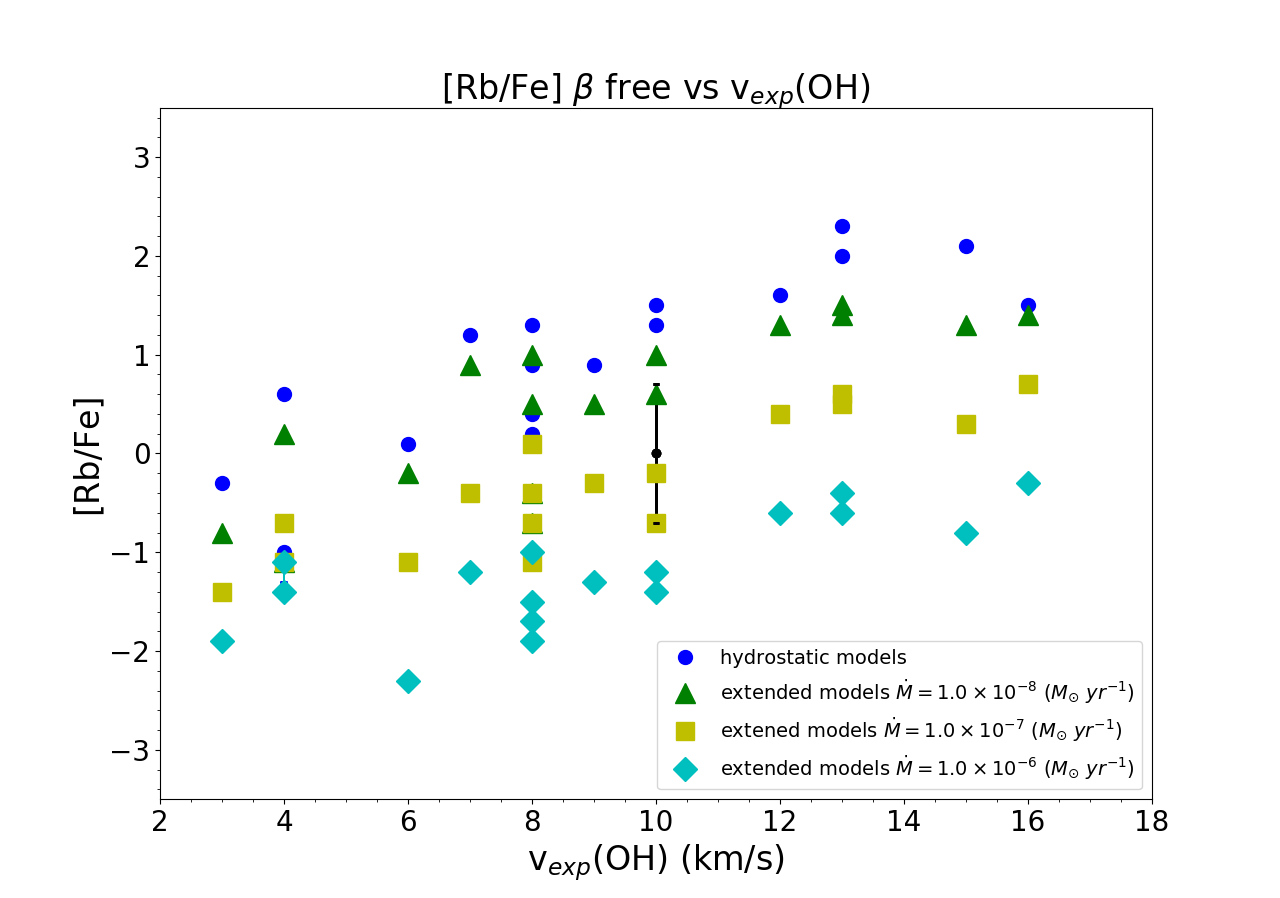}
    \caption{Rb abundances vs. the expansion velocity ($v_{exp}$(OH)) for
extended model atmospheres with $\dot{M}$ = 10$^{-8}$, 10$^{-7}$ and 10$^{-6}$
M$_{\odot}$ yr$^{-1}$ (green triangles, yellow squares and cyan diamonds, respectively) in
comparison with those obtained from hydrostatic models (blue dots).}
         \label{Rb_mass_fixed}
\end{figure}

On the other hand, Figure \ref{Rb_mass_fixed} displays the Rb results when 
$\dot{M}$ is fixed to 10$^{-8}$, 10$^{-7}$ and 10$^{-6}$ M$_{\odot}$ yr$^{-1}$. 
This could be equivalent to consider that our AGB sample stars have a similar
evolutionary stage in terms of the mass loss; of course we have a strong
degeneracy between the progenitor masses and mass loss/evolutionary stage.
In the particular case of the Li-rich AGB stars, statistical arguments
suggest that these stars should have a narrow initial mass range
\citep[see][]{dicriscienzo16}; 4-5 or 5-6 M$_{\odot}$ according to
the most recent ATON \citep{dicriscienzo16} or Monash \citep{karakaslugaro16}
AGB nucleosynthesis models, respectively. The current stellar masses from 
Table \ref{table_masses}
show, however, that there is a complicated interplay (degeneracy) between 
Li enhancement, progenitor mass and mass-loss rate and that the progenitor 
mass range of these stars may be actually broader; e.g., their current 
stellar mass and Li abundance ranges are $\sim$2.7-5.6 M$_{\odot}$ and 
$\sim$log$\varepsilon$(Li)$\sim$0.7-2.6 dex. Figure
\ref{Rb_mass_fixed} shows that the Rb abundances decrease with increasing
$\dot{M}$ and the dispersion of the Rb abundances is much lower when fixing
$\dot{M}$. The slopes (and correlation coefficients) of the Rb-$v_{exp}$(OH) 
correlations are more
similar to the one obtained with hydrostatic models. The Rb abundances from
extended models approach the hydrostatic ones with decreasing $\dot{M}$ (both
sets of Rb abundances are identical for $\dot{M}$ $\leq$ 10$^{-9}$ M$_{\odot}$
yr$^{-1}$) because the atmosphere is less extended with decreasing $\dot{M}$, as
expected. 

\begin{figure}
   \centering
   \includegraphics[width=9.55cm,angle=0]{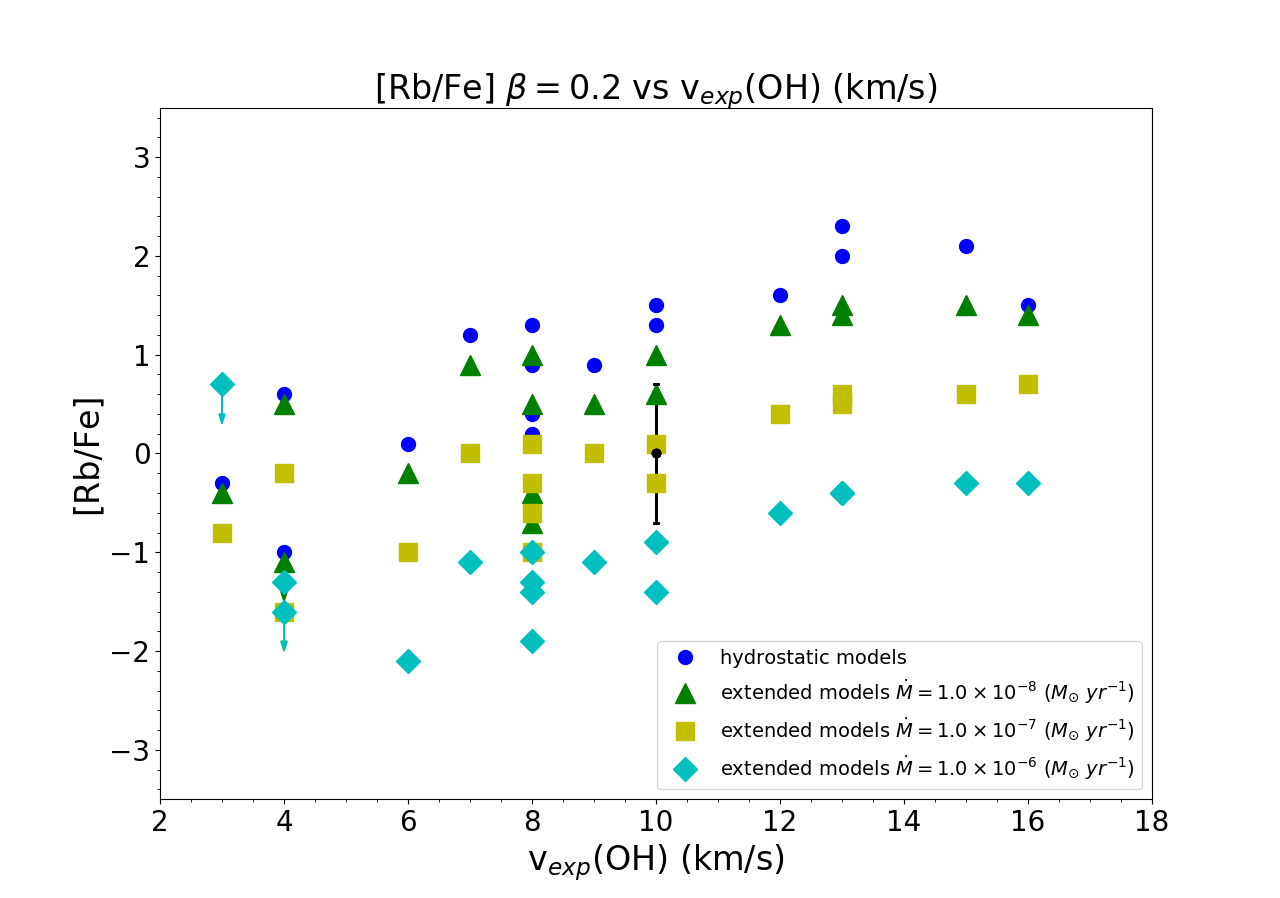}
    \caption{Rb abundances vs. the expansion velocity ($v_{exp}$(OH)) for
extended model atmospheres with $\beta$ = 0.2 and $\dot{M}$ = 10$^{-8}$, 
10$^{-7}$ and 10$^{-6}$ M$_{\odot}$ yr$^{-1}$ (green triangles, yellow 
squares and cyan diamonds, respectively) in comparison with those obtained 
from hydrostatic models (blue dots).}
		\label{Rb_beta02_mass_fixed}
\end{figure}

Finally, we fixed $\dot{M}$ and $\beta$, which could be equivalent to consider
that our AGB sample stars have the same mass loss stage and velocity profile.
Figure \ref{Rb_beta02_mass_fixed} displays the pseudo-dynamical Rb abundances versus
$v_{exp}$(OH) for $\beta$ = 0.2 and $\dot{M}$ values of 10$^{-8}$, 10$^{-7}$ and
10$^{-6}$ M$_{\odot}$ yr$^{-1}$. The Rb results when fixing both $\dot{M}$ and 
$\beta$ are very similar (with a slightly tighter correlation with $v_{exp}$(OH)) 
to those obtained when only fixing $\dot{M}$ (see Figure \ref{Rb_mass_fixed}) 
because $\beta$ = 0.2 is the most common value obtained from the best spectral 
fits (all wind parameters free); an exception is the AGB star IRAS 15193$+$3132 
(with the lowest $v_{exp}$(OH) and high $\beta$) for which only an upper limit 
to the Rb abundance ([Rb/Fe]$\leq$0.7) could be obtained because the pseudo-dynamical 
model does not converge for such unusual combination of wind parameters ($\dot{M}$=10$^{-6}$
M$_{\odot}$ yr$^{-1}$; $\beta$=0.2; $v_{exp}$(OH)=3 km s$^{-1}$) coupled with 
[Rb/Fe]$<$0.7 dex (see Figure \ref{Rb_beta02_mass_fixed}).

\begin{figure}
   \centering
   \includegraphics[width=9.1cm,angle=0]{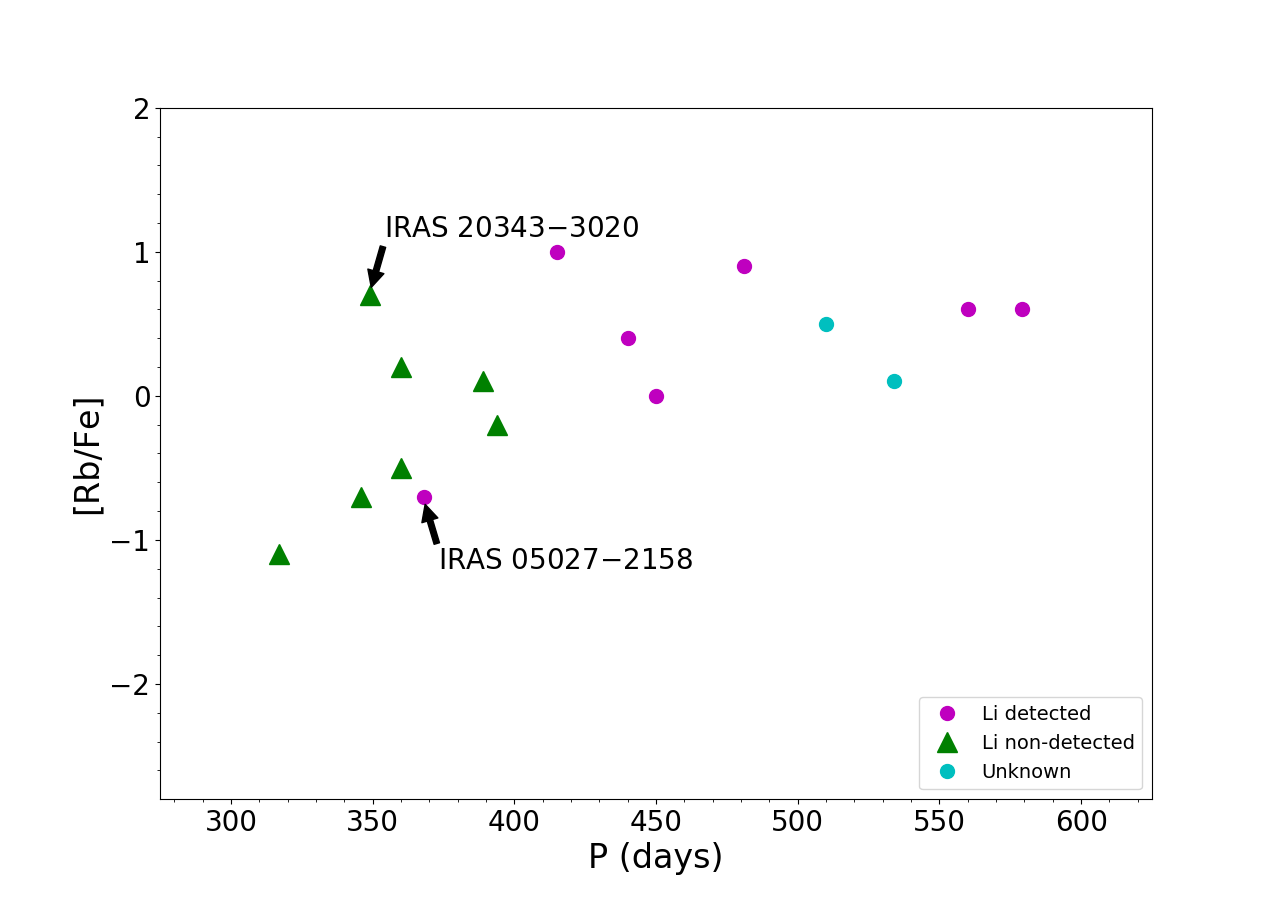}
   \caption{[Rb/Fe] pseudo-dynamical abundances versus variability period (\textit{P}).
The Li-rich and Li-poor stars are marked with magenta dots and green triangles, respectively.
The two stars where Li could not be estimated are marked with cyan dots (see
text).}
         \label{rb_vs_period}
   \end{figure}

Figure \ref{rb_vs_period} displays the [Rb/Fe] abundances from the best spectral
fits versus the variability periods \textit{P}. As already mentioned, our sample is
composed by AGB stars of different progenitor masses and evolutionary stages.
Most stars with \textit{P} $>$ 400 days are Li-rich and present some Rb
enhancement\footnote{The only exceptions are IRAS 04404$-$7427 and IRAS
19059$-$2219, whose optical counterparts are too red to estimate their Li
abundances \citep[i.e., the S/N at 6708 \AA~ is too low;
see][]{garcia-hernandez07}}, which suggests that, on average, these stars are
more massive stars experiencing HBB and/or more evolved stars (because of the
longer periods) than the group of non Li-rich (and generally Rb-poor) stars with
\textit{P} $<$ 400 days. The stars IRAS 05027$-$2158 (P$=$368 days) and IRAS
20343$-$3020 (P$=$349 days) are exceptions in the latter group. IRAS 05027$-$2158
is slightly Li-rich and Rb-poor, suggesting that it is a relatively massive AGB
star (say $\sim$3.5$-$4.5 M$_{\odot}$\footnote{The initial mass for HBB
activation is model dependent; i.e., at solar metallicity HBB is activated at
$\sim$3.5 and 4.5 M$_{\odot}$ depending on the mass-loss and convection
prescriptions used in the models \citep[see e.g.][for more
details]{garcia-hernandez13}.}) at the beginning of the TP phase (e.g., in an
inter-pulse period just before or after the super Li-rich phase) but it is not
evolved enough for efficient Rb production \citep[see][]{garcia-hernandez13}. On
the other hand, IRAS 20343$-$3020 is slightly Rb-rich and Li-poor, which
suggests a more advanced evolutionary stage and a slightly higher initial mass
(say $\sim$4.0$-$5 M$_{\odot}$) than IRAS 05027$-$2158 \citep[see Fig. 1
in][]{garcia-hernandez13}. 

\section{Comparison with AGB nucleosynthesis models} \label{comparison_section}

\begin{figure*}
   \centering
   \includegraphics[width=9.1cm,angle=0]{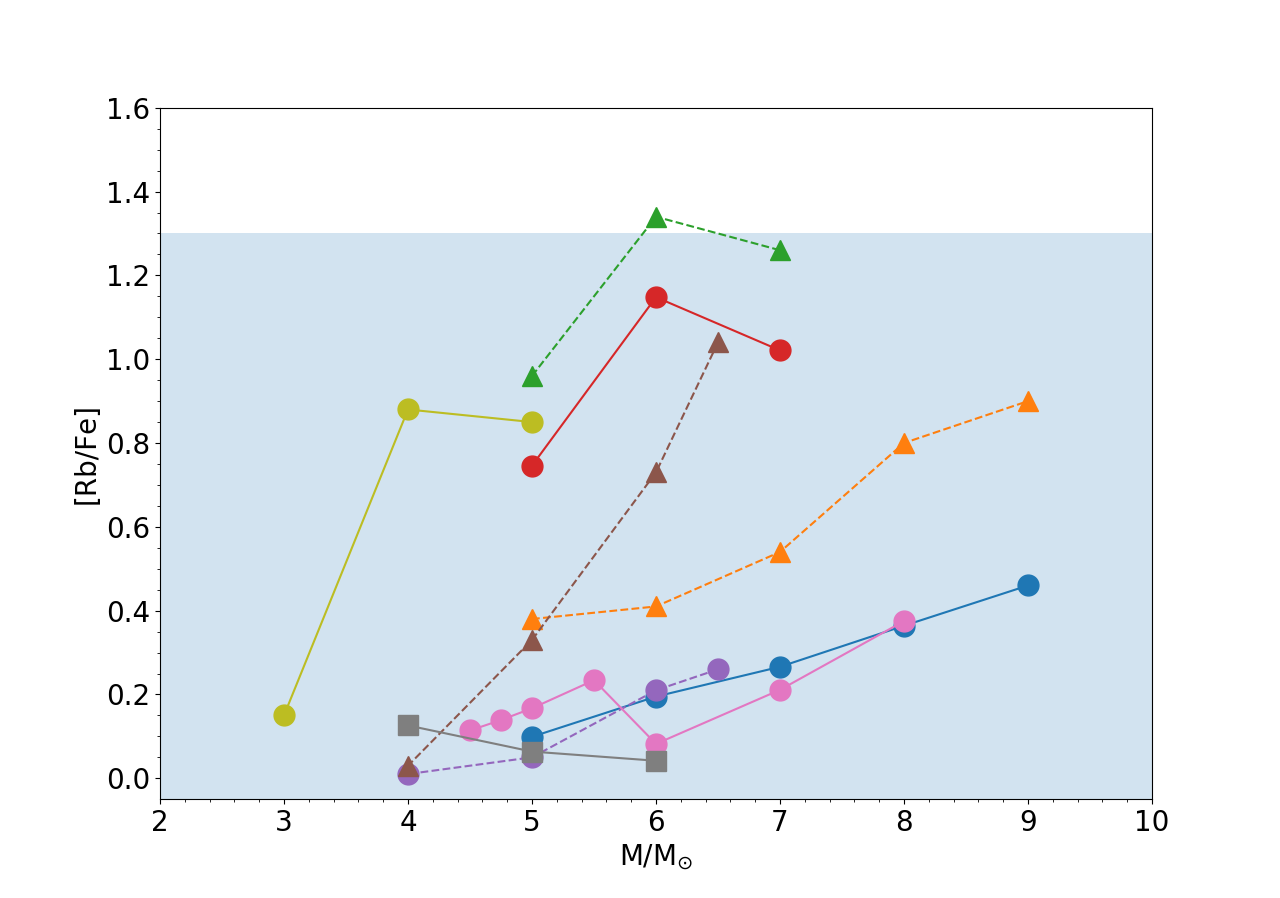}
    \includegraphics[width=9.1cm,angle=0]{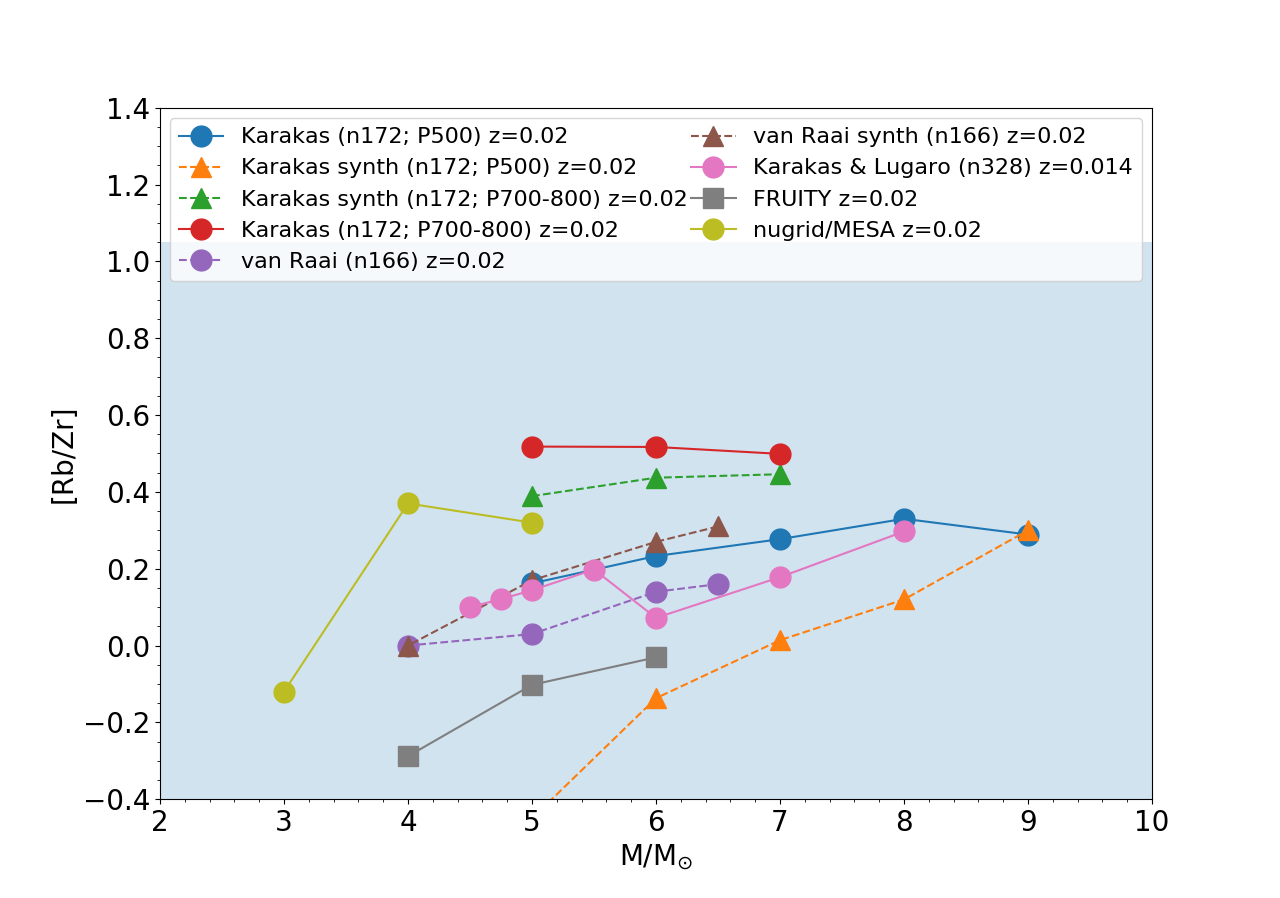} 
      \caption{Model predictions from \cite{vanraai12}, \cite{karakas12}, 
      \cite{karakaslugaro16}, FRUITY database and \cite{pignatari16}: stellar 
      mass vs. [Rb/Fe] (\textit{left panel}) and [Rb/Zr] (\textit{right panel}). 
      The abundances from the last computed thermal pulse and from the synthetic 
      evolution calculations are shown with dots and triangles, respectively. 
      \textit{P} represents the period for the beginning of the superwind phase 
      and \textit{n} is the number of species in the nucleosynthesis network. The 
      shaded regions mark the range of the new Rb abundances and [Rb/Zr] ratios 
      obtained in our sample with extended models.}
         \label{models_RbZr_M}
   \end{figure*}

In Figure \ref{models_RbZr_M} we compare our new [Rb/Fe] abundances and [Rb/Zr] 
ratios with solar metallicity massive (3$-$9 M$_{\odot}$) AGB predictions from
several nucleosynthesis models: \cite{vanraai12}, \cite{karakas12},
\cite{karakaslugaro16} (Monash), \cite{pignatari16} (NuGrid/MESA)
and \cite{cristallo15} (FRUITY\footnote{FUll-Network Repository of
Updated Isotopic Tables and Yields: http:// fruity.oa-teramo.inaf.it/.}). The
predicted [Rb/Fe] abundances and [Rb/Zr] ratios ranges are 0.00$-$1.35 and
$-$0.45$-$0.52 dex, respectively.

The Monash models \citep{vanraai12, karakas12, karakaslugaro16} use the stellar 
evolutionary sequences calculated with the Monash version of the
Mount Stromlo Stellar Structure Program \citep{frost96}, which uses 
the \cite{vassiliadis93}
mass-loss prescription on the AGB. A post-processing code is used to obtain 
in detail the nucleosynthesis of a large number of species, including the 
\textit{s}-process abundances. Due to convergence difficulties, the stellar
evolution models used in the calculations  are not always evolved until the end of the
superwind phase and synthetic models have been used to estimate the effect of 
remaining TPs and to completely remove the envelope. We refer the reader to
\cite{vanraai12}, \cite{karakas12} and \cite{karakaslugaro16} for more details
about the theoretical models. Here, we only report the main differences between these 
models and that are basically the following: i) the use of different nuclear 
networks; i.e., the
total number of nuclear species considered and the values of some reaction rates and
neutron-capture cross sections (see below); and ii) the use by \cite{karakas12} of a
modified \cite{vassiliadis93} mass-loss prescription, which delays the
beginning of the superwind phase until the pulsation period reaches values of
700-800 days (instead of the value of 500 days used in the other models), resulting 
in a higher Rb production. 

The NuGrid/MESA and FRUITY models assume AGB mass-loss
prescriptions, nuclear physics inputs and treatments of convection
different from the Monash models. In particular, the \citet{blocker95} and
\citet{straniero06} mass-loss formulae for the AGB phase are assumed by the
NuGrid/MESA and FRUITY models, respectively. Furthermore, these models produce 
self-consistently the $^{13}$C neutron source as a result of the different 
convective boundary mixing scheme and treatments of the convective borders, 
while in the Monash models the  mixing required to produce the $^{13}$C neutron 
source is included in a parametrized way during the post processing and it is 
typically not included in massive AGB stars, following theoretical \citep{gorielysiess04} 
and observational indications \citep{garcia-hernandez13}
\citep[see also][for more details]{pignatari16,cristallo15,karakaslugaro16}. In relation 
to the main results: i) the NuGrid/MESA solar metallicity massive AGB models are 
qualitatively similar to the Monash models in terms of HBB and light \textit{s}-process 
element production (of the elements from Rb to Zr) are seen at the stellar surface, 
the latter due to the activation of the $^{22}$Ne neutron source and the subsequent 
operation of the TDU; and ii) the FRUITY solar metallicity 
massive AGB models are different to the Monash and NuGrid/MESA models because 
these models experience very inefficient TDU, hence the signature of the nucleosynthesis 
due to the $^{22}$Ne neutron source is not visible at the stellar surface.

Figure \ref{models_RbZr_M} shows that the FRUITY massive AGB models predicts final 
[Rb/Fe] $<$ 0.15, which does not explain the observed range of Rb abundances and 
[Rb/Zr] ratios; specifically the [Rb/Zr] ratios remains negative for all masses. 
Another difference of the FRUITY models with the Monash and NuGrid models 
is that the FRUITY models do not predict HBB to occur in AGB stars, unless the metallicity 
is very low, at least ten times lower than solar. However, spectroscopic 
observations of massive AGB stars demonstrate that they experience HBB; as evidenced by: i) 
strong Li overabundances observed in massive AGB stars in the Galaxy
\citep[Fe/H=0.0; e.g.,][]{garcia-hernandez07,garcia-hernandez13}, the
Magellanic Clouds \citep[Fe/H=$-$0.7-$-$0.3; e.g.,][]{plez93,smith95,garcia-hernandez09} 
and the dwarf galaxy IC 1613
\citep[Fe/H=$-$1.6; e.g.,][]{menzies15}; ii) N enhancements and low
$^{12}$C/$^{13}$C ratios in Magellanic Cloud Li-rich massive AGBs
\citep[e.g.,][]{plez93,mcsaveney07}.
The lack of HBB in the FRUITY predictions is also at
odds with the observations of the so-called type I planetary nebulae in very
different metallicity environments and galaxies; which are expected to be the
descendants of HBB massive AGB stars based on their strong N and He
overabundances \citep[see e.g.][and references therein]{stanghellini06,
karakas09,leisy96,garcia-rojas16}.

It is to be noted here that the several Monash AGB models
\citep{vanraai12,karakas12,karakaslugaro16} mentioned above notably use different 
rates for the $^{22}$Ne($\alpha,n$)$^{25}$Mg reaction, which drives the production
of \textit{s}-process elements in massive AGB stars. In particular, 
\cite{karakaslugaro16} use the  $^{22}$Ne($\alpha, n$)$^{25}$Mg reaction from 
\cite{iliadis10}, neutron-capture cross section of the Zr isotopes \citep{lugaro14}, and a
more extended nuclear network of 328 species (from H to S, and then from Fe to
Bi). The \cite{vanraai12} models, instead, use a nuclear network of 166 species (up to
Nb) and the $^{22}$Ne($\alpha, n$)$^{25}$Mg reaction rate from \cite{karakas06},
while \citet{karakas12} explored different networks (166, 172 and 320 species)
and $^{22}$Ne($\alpha, n$)$^{25}$Mg reaction rates; from \citet{karakas06},
\cite{iliadis10} and \cite{angulo99}, NACRE.

The \cite{vanraai12} models (from 4 to 6.5 M$_{\odot}$ at $Z$ = 0.02; Fig.
\ref{models_RbZr_M}) show that both the [Rb/Fe] abundances and [Rb/Zr] ratios
increase with the initial mass of the AGB star, as the star becomes hotter and 
the $^{22}$Ne($\alpha, n$)$^{25}$Mg reaction is more efficiently activated. However, 
the [Rb/Fe] abundances from the last computed TP are too low (ranging from 0.0 to 
0.26 dex). The corresponding Rb abundances ([Rb/Fe]$\sim$0.0$-$1.0 dex) from the
synthetic evolution calculations cover most of the Rb abundances observed;
although they cannot explain the star IRAS 05151$+$6312 with [Rb/Fe]=1.3 dex.
Such high Rb abundances can be reached by the synthetic calculations of the
solar metallicity 6 and 7 M$_{\odot}$ AGB models with delayed superwinds of
\cite{karakas12} when using the faster NACRE rate for the $^{22}$Ne($\alpha,
n$)$^{25}$Mg reaction.
Finally, the \cite{karakaslugaro16}
models (from 4.5 to 8 M$_{\odot}$ at $Z$ = 0.014\footnote{According to the more
recent solar abundances from \citealt{asplund09}}; Fig. \ref{models_RbZr_M})
predict lower Rb abundances than the \cite{karakas12} models of the same mass
and similar metallicity, mostly due to the implementation of the delayed superwind 
and the use of the NACRE rate in \cite{karakas12}.

The NuGrid/MESA models (from 3 to 5 M$_{\odot}$ at $Z$ = 0.02; Fig. 
\ref{models_RbZr_M}) reproduce the observed [Rb/Fe] and [Rb/Zr] ranges,
up to 0.9 and 0.4 dex, respectively. However, we note that only in the 5
M$_{\odot}$ case the NuGrid/MESA models see signature of HBB and predict O-rich stars. 
The 3 and 4 M$_{\odot}$ cases become C-rich stars and do not experience HBB, which
is at odds with our sample of O-rich stars \citep{garcia-hernandez06}.

Regarding the [Rb/Zr] ratios, obviously also in this case
the higher [Rb/Zr] ratios are obtained from the models
with delayed superwind (P = 700-800), however, these [Rb/Zr] ratios are still lower than our
observed values. The maximum value from the AGB models is [Rb/Zr] = 0.52 for 
M = 5 M$_{\odot}$, and the maximum value from our observations is [Rb/Zr] = 1.05.
A possible explanation is that Zr could be depleted
into dust \citep[see e.g.][]{vanraai12, zamora14}, producing the differences between
the theoretical and observational [Rb/Zr] ratios. Abundance measurements of similar 
\textit{s}-elements such as Sr and Y would be needed in order to clarify this problem.

\section{Conclusions}

We have reported new Rb and Zr abundances determined from the 7800 $\AA$ Rb I line and
the 6474 $\AA$ ZrO bandhead, respectively, in a complete sample of massive 
Galactic AGB stars, previously studied  with hydrostatic models, using more 
realistic extended atmosphere models and a modified version of the spectral synthesis 
code Turbospectrum, which considers the presence of a circumstellar envelope 
with a radial wind. The Rb abundances are much lower (in some cases even 1-2 dex) 
with the pseudo-dynamical models, while the Zr abundances are close to the 
hydrostatic ones because the 6474 \AA~ZrO bandhead is formed deeper in the atmosphere and 
is less affected than the 7800 \AA~Rb I resonante line by the circumstellar effects.

We have studied the sensitivity of 
the determined abundances to variations in the stellar (T$_{eff}$) and wind 
($\dot{M}$, $\beta$ and $v_{exp}$) parameters. The Rb abundances are very 
sensitive to the mass loss rate $\dot{M}$ but much less to the $\beta$ 
parameter and $v_{exp}$(OH). The 
Zr abundances, instead, are not affected by variations of the stellar and wind 
parameters. The Rb abundances from extended models are lower than those obtained from the
hydrostatic ones, and the difference is larger in the stars with the highest
Rb abundances in the hydrostatic case. We have represented the hydrostatic 
and pseudo-dynamical Rb abundances against the $v_{exp}$(OH), which can be used as a
mass indicator independent of the distance, and we have observed 
a flatter correlation. The difference between the hydrostatic and
pseudo-dynamical Rb abundances increases with increasing the $v_{exp}$(OH),
due to the fact that the presence of a circumstellar envelope affects more strongly the
more massive stars. Furthermore, the dispersion of the correlation between the Rb abundance
and $v_{exp}$(OH) is larger in the pseudo-dynamical case. When we fix the wind parameters
$\dot{M}$ (i.e., equivalent to assuming that our AGB sample stars have a similar
evolutionary stage in terms of mass loss), and/or $\beta$ (the same velocity profile)
the dispersion is lower.

The Monash nucleosynthesis theoretical predictions reproduce the range of new 
Rb and Zr abundances although [Rb/Fe] values above 1.0 can be matched only if 
the superwind is delayed to after the period reaches 700-800 days. We also note 
that the rate of the $^{22}$Ne($\alpha,n$)$^{25}$Mg  reaction is crucial, but 
still hampered by large systematic uncertainties \citep[see e.g.][]{bisterzo16, massimi17}.
Underground measurements, planned, e.g., at LNGS-LUNA (Laboratory for Underground 
Nuclear Astrophysics) will help to resolve the current issues. The FRUITY 
massive AGB models predict Rb abundances much lower than observed and negative
[Rb/Zr] ratios, at odds with the observations. The NuGrid/MESA models of 4 and 
5 M$_{\odot}$ predict [Rb/Fe] as high as 0.9 dex, however, the 4M$_{\odot}$  model 
do not experience HBB and becomes C-rich, while our sample stars are clearly O-rich. The
maximum observed [Rb/Zr] ratios are still more than a factor of two larger than predicted
by the nucleosynthesis models. A possible explanation to this difference 
between the observations
and the predictions is that Zr could be depleted into dust. Observations of other 
\textit{s}-process elements Sr and Y belonging to the same first peak as Rb and Zr will 
help clarifying this mismatch.

In summary, the [Rb/Fe] abundances and [Rb/Zr] ratios previously 
derived with hydrostatic models are certainly not
predicted by the most recent theoretical models of AGB nucleosynthesis. In
particular, the highest [Rb/Fe] abundances and [Rb/Zr] ratios observed in
massive Galactic AGBs are much larger than theoretically predicted.  The new
[Rb/Fe] abundances and [Rb/Zr] ratios as obtained from our simple (but more 
realistic) pseudo-dynamical
model atmospheres are much lower in much better agreement with the theoretical
predictions, significantly resolving the mismatch between the observations and the
nucleosynthesis models in the more massive AGB stars. This confirms the earlier
Zamora et al. preliminary results on a smaller sample of massive O-rich AGB
stars but here we find that the Rb abundances are strongly dependent of the wind
mass-loss $\dot{M}$, which is basically unknown in our AGB stars sample.
Follow-up radio observations (e.g. the rotational lines of the several CO
isotopologues) of these massive Galactic AGB stars are encouraged in order to
get precise mass-loss rates estimates, which are needed to break the actual
models degeneracy and obtain more reliable (no model-dependent) Rb abundances in
massive AGB stars.


\begin{acknowledgements}
This work is based on observations at the 4.2 m William Herschel Telescope
operated on the island of La Palma by the Isaac Newton Group in the
Spanish Observatorio del Roque de Los Muchachos of the Instituto de
Astrofisica de Canarias. Also based on observations with the ESO 3.6 m
telescope at La Silla Observatory (Chile). We thank Marco Pignatari and 
Umberto Battino for providing information about the Nugrid/MESA models.
V.P.M. acknowledges the financial support from the Spanish Ministry of Economy
and Competitiveness (MINECO) under the 2011 Severo Ochoa Program MINECO
SEV-2011-0187. D.A.G.H. was funded by the Ram\'on y Cajal fellowship number
RYC-2013-14182. V.P.M., O.Z., D.A.G.H. and A.M. acknowledge support provided by
the MINECO under grant AYA-2014-58082-P. M.L. is a Momentum
(“Lend\"ulet-2014” Programme) project leader of the Hungarian Academy of
Sciences. M. L. acknowledges the Instituto de Astrofísica de Canarias for
inviting her as a Severo Ochoa visitor during 2015 August when part of this work
was done. This paper made use of the IAC Supercomputing facility HTCondor
(http://research.cs.wisc.edu/htcondor/), partly financed by the Ministry of
Economy and Competitiveness with FEDER funds, code IACA13-3E-2493. This work
benefited from discussions at The 12th Torino Workshop on Asymptotic Giant
Branch Stars in August 2016 supported by the National Science Foundation under
Grant No. PHY-1430152 (JINA Center for the Evolution of the Elements).
\end{acknowledgements}

\begin{appendix}

\section{Complete sample}\label{Append_sample}
\begin{figure*}
   \centering
   \includegraphics[width=9.1cm,height=6.5cm,angle=0]{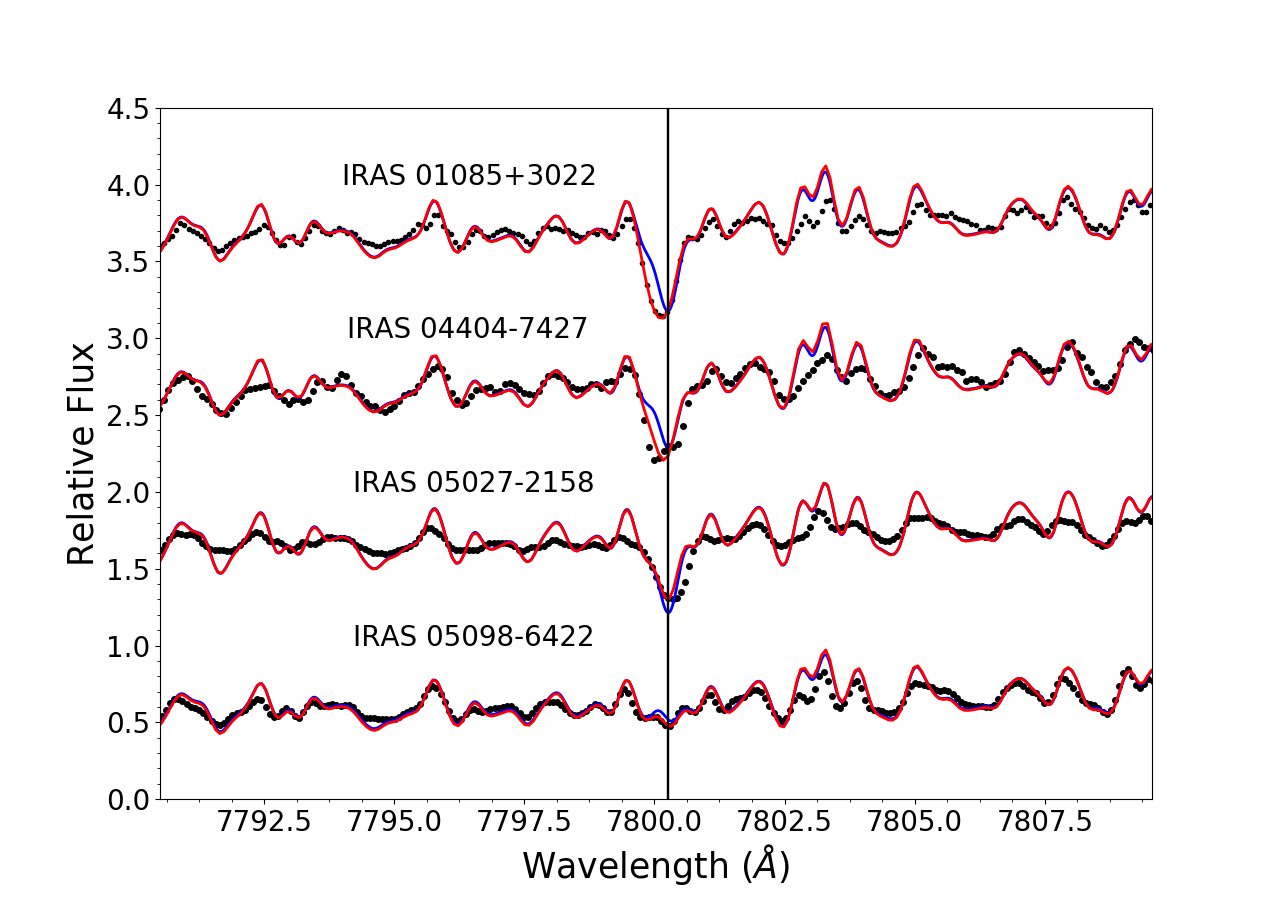}
   \includegraphics[width=9.1cm,height=6.5cm,angle=0]{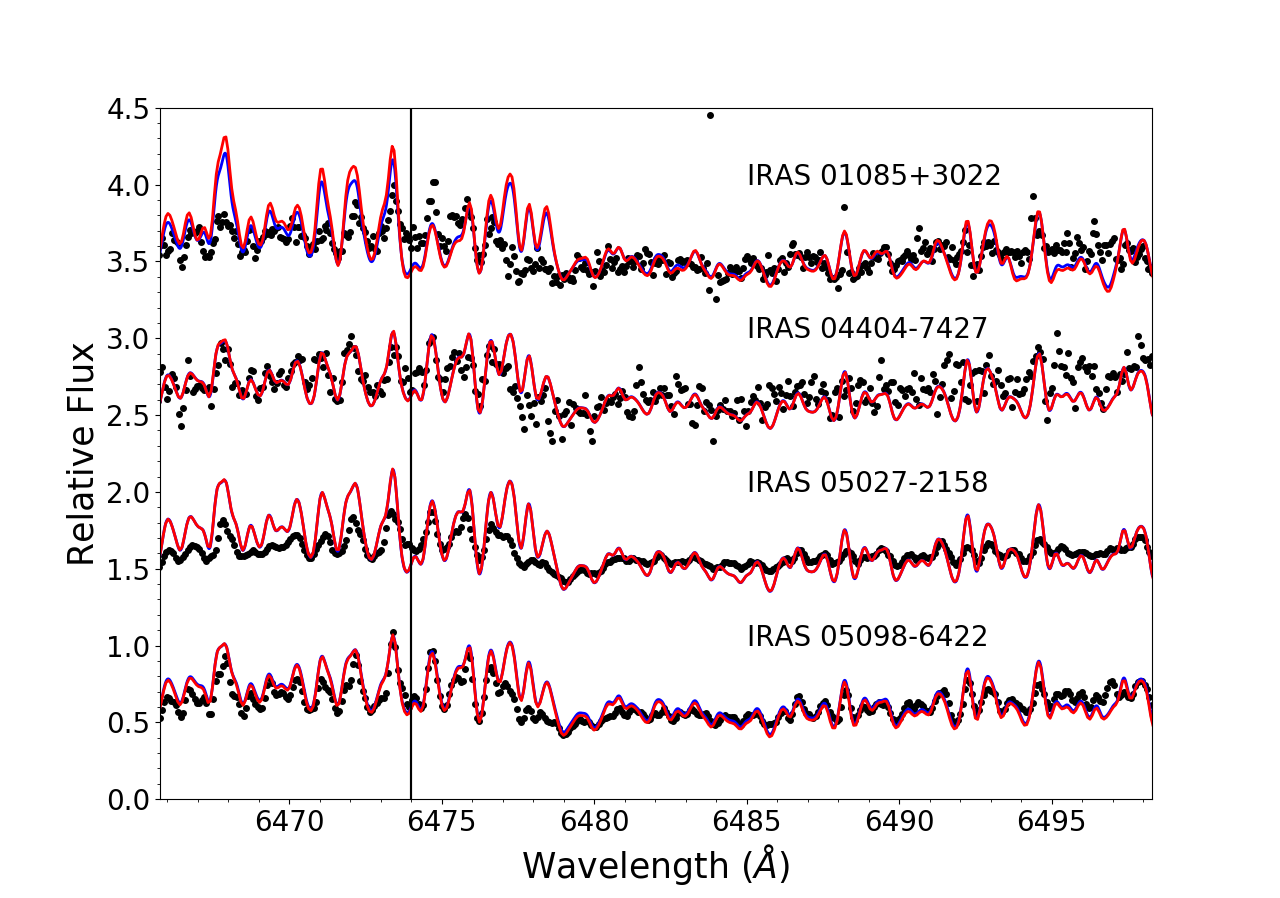} 
   \includegraphics[width=9.1cm,height=6.5cm,angle=0]{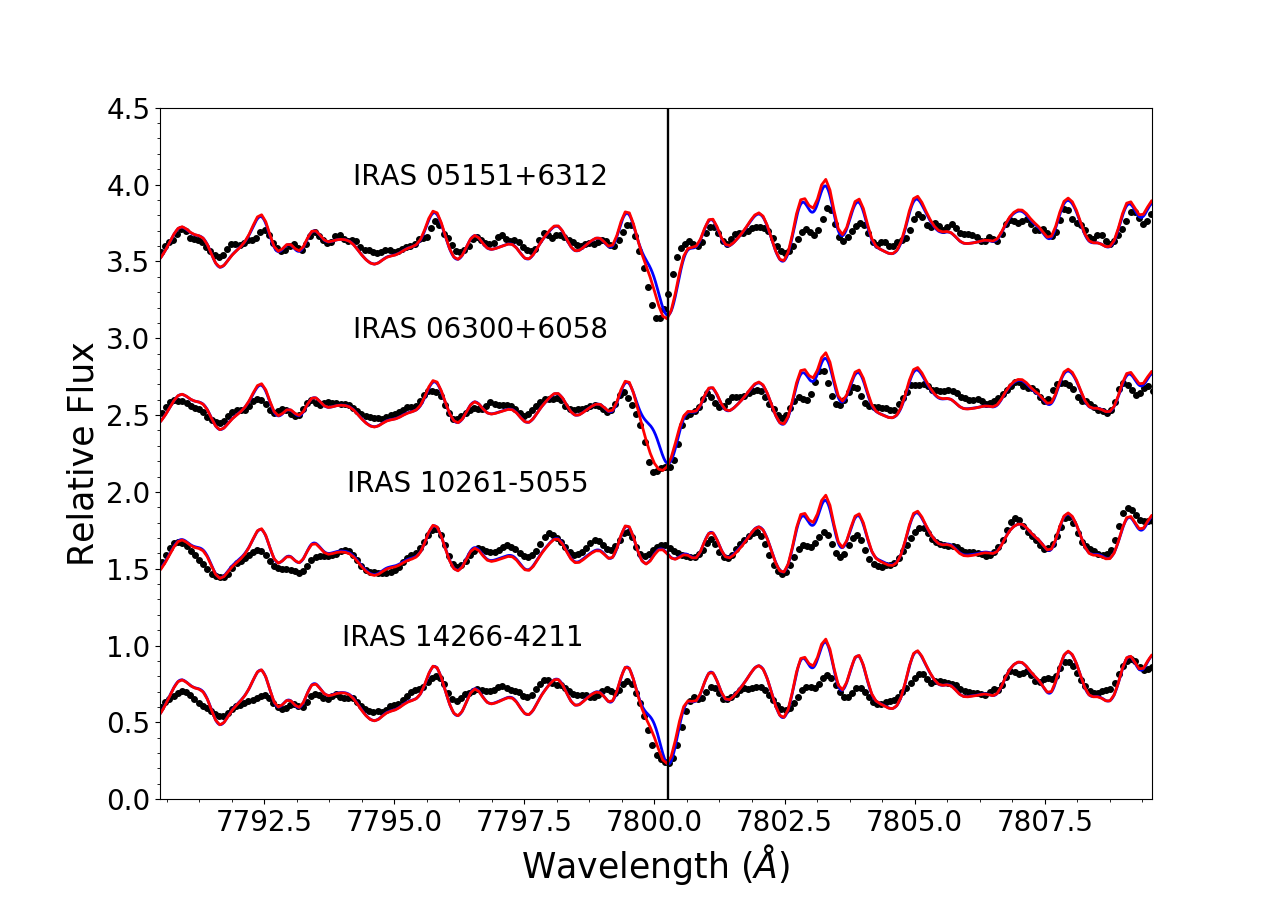}
   \includegraphics[width=9.1cm,height=6.5cm,angle=0]{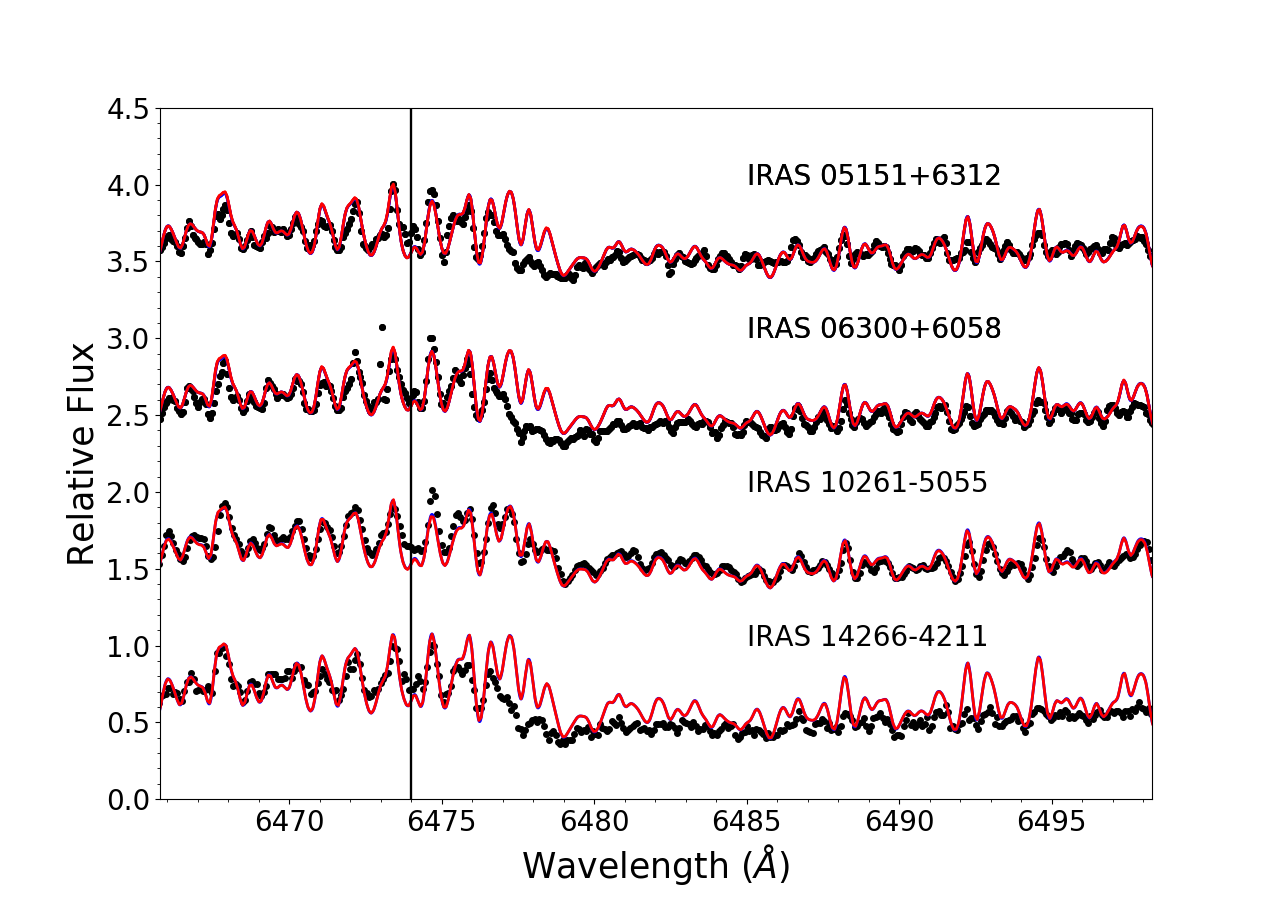}
   \includegraphics[width=9.1cm,height=6.5cm,angle=0]{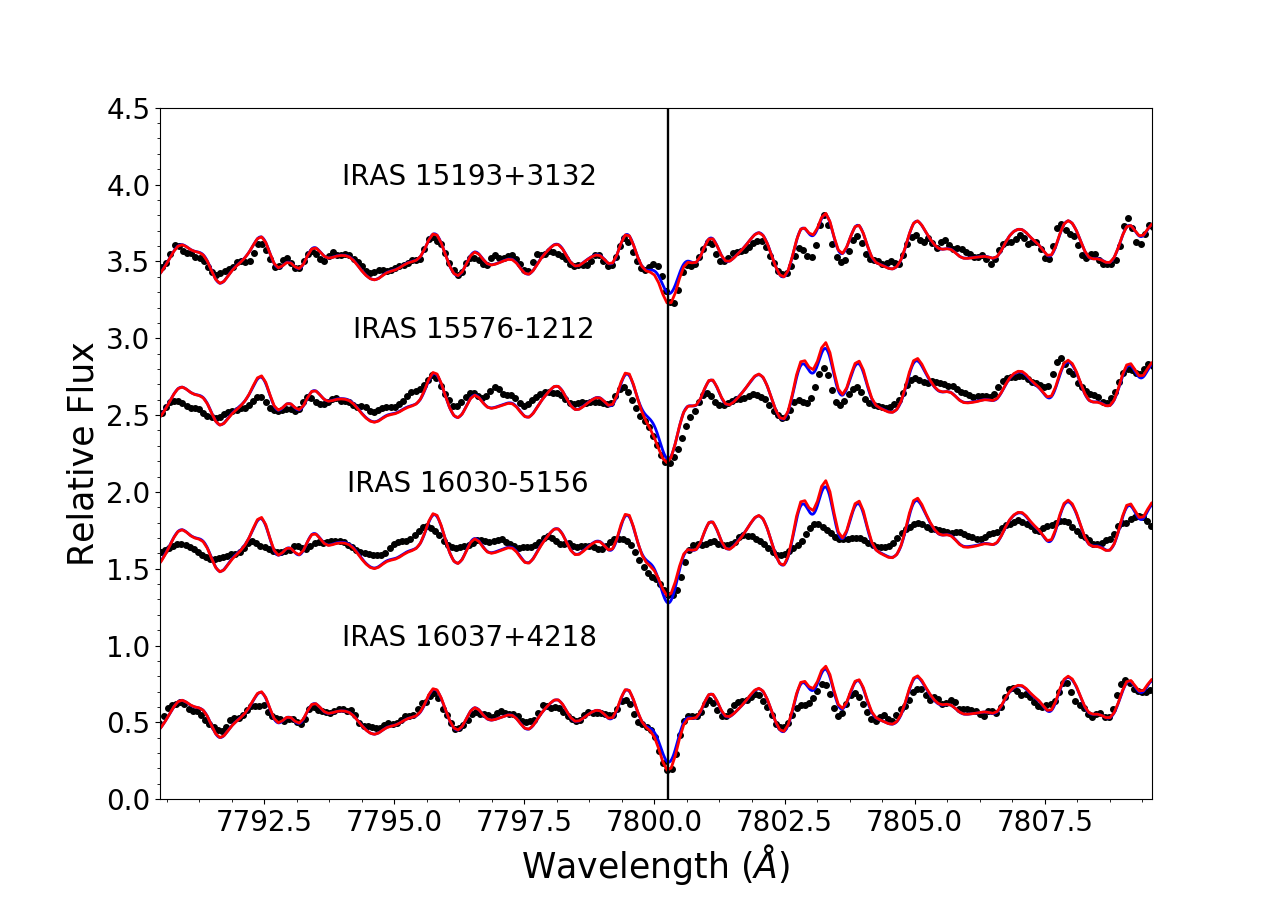}
   \includegraphics[width=9.1cm,height=6.5cm,angle=0]{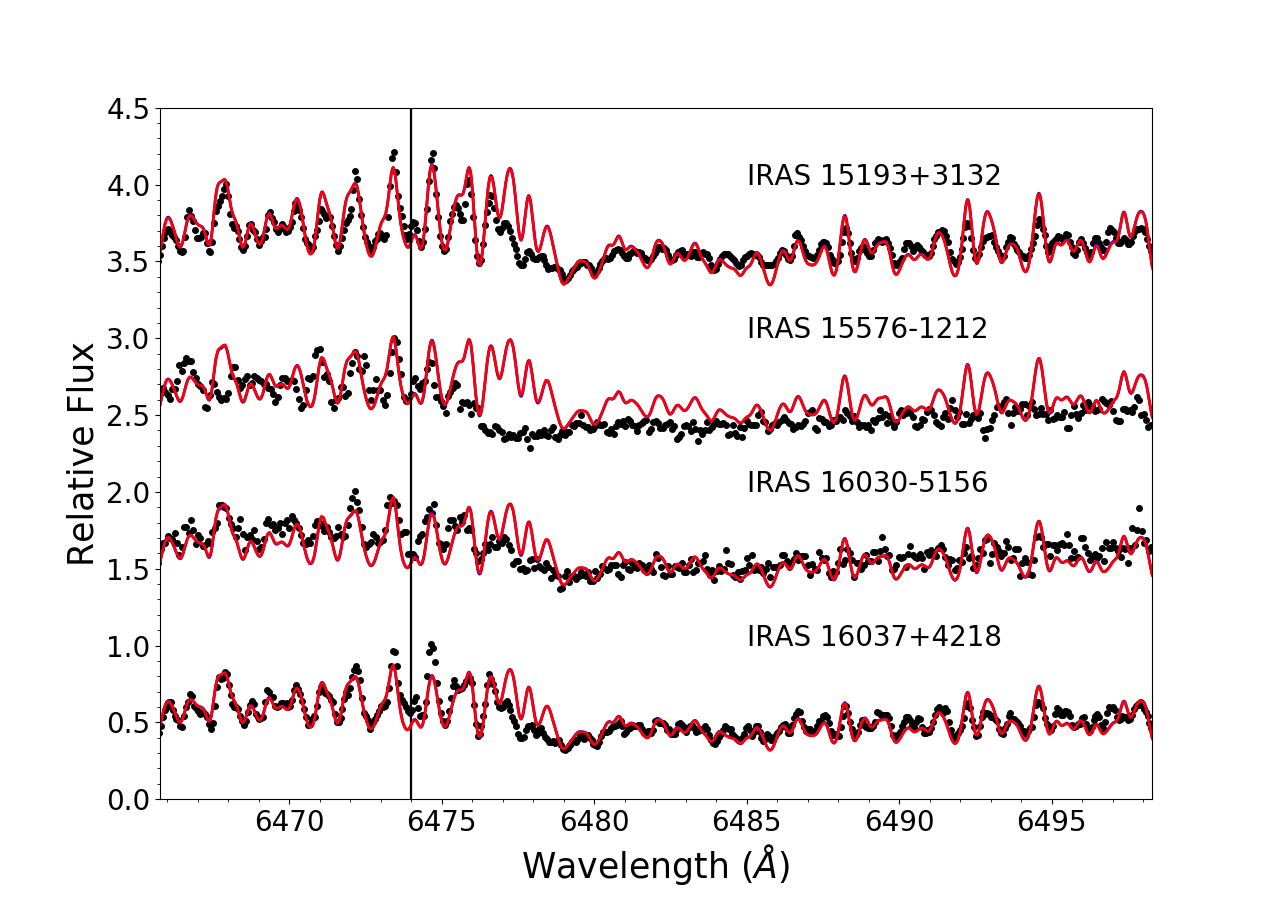}
   \includegraphics[width=9.1cm,height=6.5cm,angle=0]{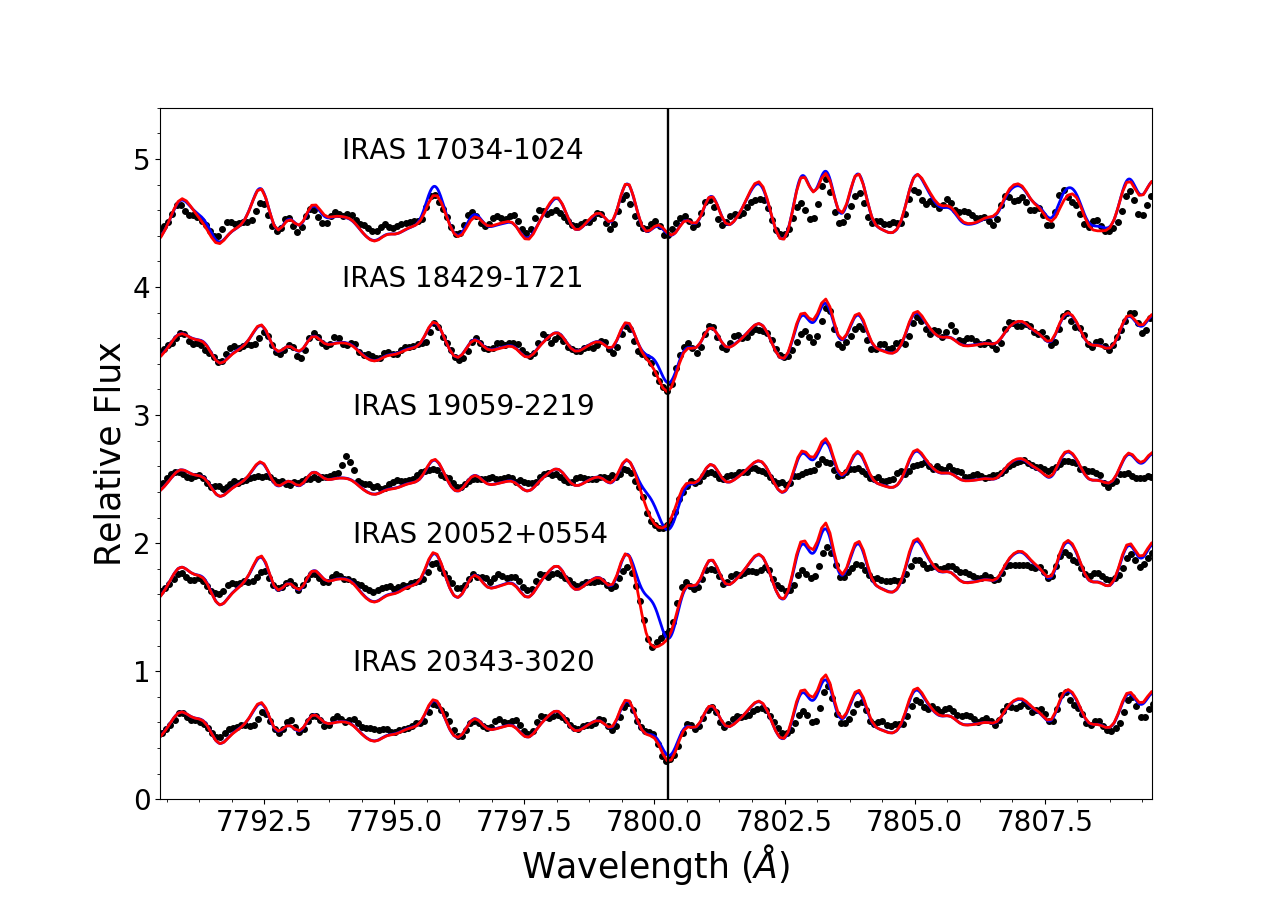}
   \includegraphics[width=9.1cm,height=6.5cm,angle=0]{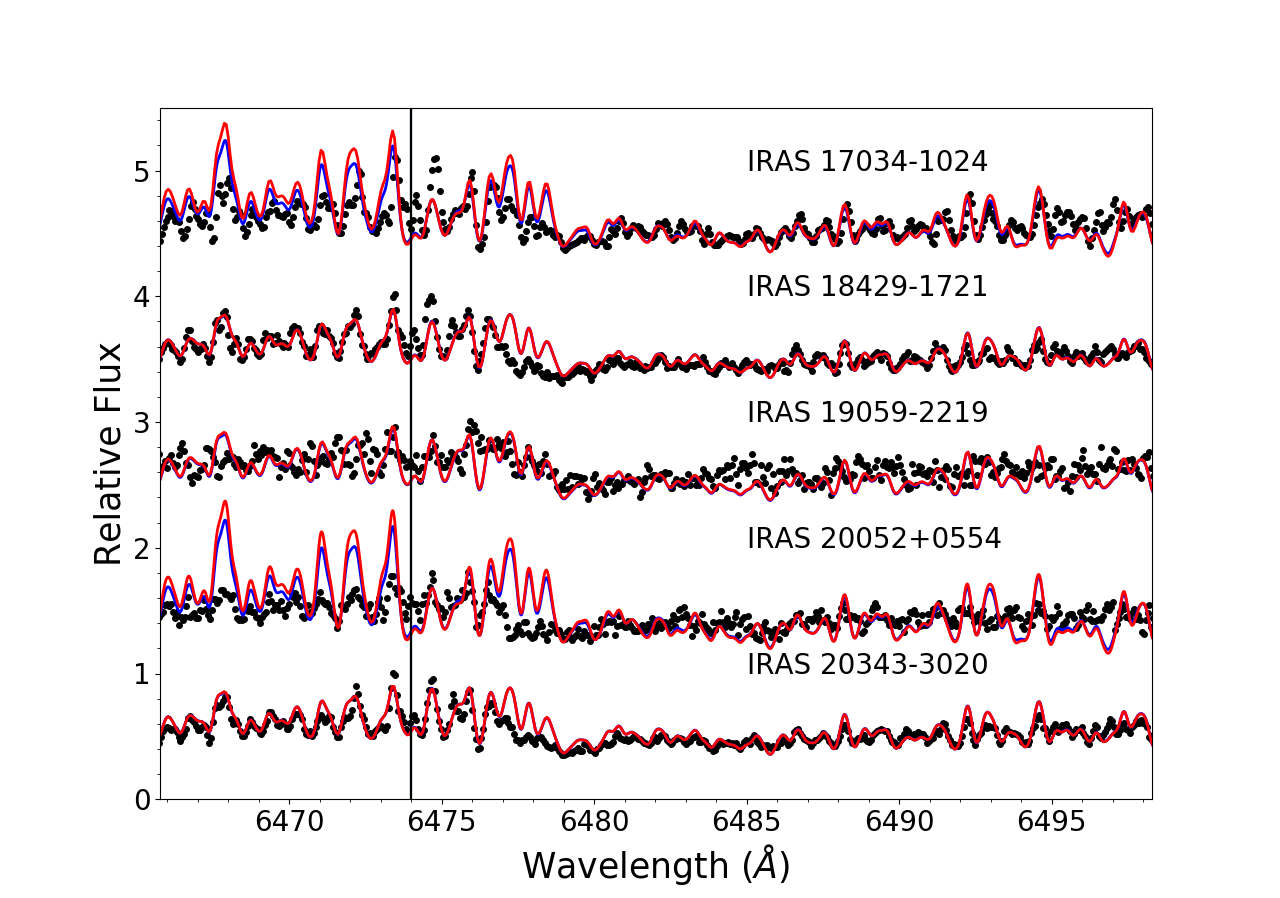}
      \caption{Observed spectra (black dots), best hydrostatic (blue lines) and 
      pseudo-dynamical (red lines) fits of our sample of AGB stars in the regions 
      of 7800 $\AA$ Rb I line (\textit{left panels}) and 6474 $\AA$ ZrO bandhead 
      (\textit{right panels}). The parameters of the best fits model atmospheres 
      are indicated in Table \ref{table_beta_free}. The plots are displayed in 
      increasing R.A. order.}
\end{figure*}

\end{appendix}

\end{document}